\documentclass[onecolumn,noshowpacs,nofootinbib,colorlinks,hyperindex,unicode,11pt]{revtex4}
\usepackage{graphicx}
\usepackage{tikz}

\usepackage[compat=1.1.0]{tikz-feynman}

\usepackage{amsfonts,amsmath,amssymb,tikz,upgreek,accents}
\usepackage{indentfirst}
\usepackage{hyperref,upgreek}
\usepackage{bm,graphics,color}
\usepackage{feynmp-auto}
\usepackage{slashed}
\counterwithout{equation}{section}
\usepackage{hyperref}

\hypersetup{hidelinks,backref=true,pagebackref=true,hyperindex=true,colorlinks=true,breaklinks=true,urlcolor= blue}
\hypersetup{%
  colorlinks = true,
  linkcolor  = blue,
  citecolor = cyan,
}
\usepackage{textgreek}
\usepackage{color,xcolor}
\newcommand{\gdualn}[1]{\overset{\:{}^{{}^{\boldsymbol{\neg}}}}{\smash[t]{#1}}} 

\def\df{\mbox{$\displaystyle\gdualn{\mathfrak{f}}$}}

\def\0{\mbox{\boldmath$\displaystyle\mathbb{O}$}}

\def\I{\openone}
\def\openone{\mathbb I}
\def\s{\mbox{\boldmath$\displaystyle\boldsymbol{\sigma}$}}

\def\x{\mbox{\boldmath$\displaystyle\boldsymbol{x}$}}
\def\p{\mbox{$\displaystyle\bf{p}$}}

\newcommand\orcidroldao{{\href{https://orcid.org/0000-0003-3978-532X}{\orcidicon}}}
\newcommand{\orcidicon}{%
	\begin{tikzpicture}
	\draw[lime, fill=lime] (0,0)
		circle [radius=0.16]
		node[white] {{\fontfamily{qag}\selectfont \tiny ID}};
	\draw[white, fill=white] (-0.0625,0.095)
		circle [radius=0.007];
	\end{tikzpicture}	\hspace{-2mm}
}

\def\nn{\nonumber }
\usepackage{float,xcolor,upgreek}
\usepackage{color} 
\usepackage{tikz-feynman}
\tikzfeynmanset{compat=1.0.0}

\newcommand{\beq}{\begin{eqnarray}}
\newcommand{\eeq}{\end{eqnarray}}
\newcommand{\bea}{\begin{eqnarray}}
\newcommand{\eea}{\end{eqnarray}}

\begin{document}

\title{Fermionic dark matter interaction with the photon and the proton in the quantum field-theoretical approach and generalizations}

\author{G. B. de Gracia}
\affiliation{Federal University of ABC, Center of Mathematics,  Santo Andr\'e, 09210-580, Brazil.}
\email{gabriel.gracia@ufabc.edu.br}
\email{roldao.rocha@ufabc.edu.br}
\author{R. da Rocha\orcidroldao\!\!}
\affiliation{Federal University of ABC, Center of Mathematics,  Santo Andr\'e, 09210-580, Brazil.}

\begin{abstract}
Mass dimension one quantum fields,  constructed upon eigenspinors of the
charge conjugation operator with dual
helicity (ELKO), are prime candidates to 
describe dark matter. Their interaction with the photon and the proton, in scattering processes, are here explored and 
discussed, using both the standard QFT with 
a Maxwell gauge field and a generalized QFT 
also involving a Podolsky gauge sector. 
Renormalization and radiative corrections 
are analyzed in the ELKO setup, in the 
context of a generalized spinor dual 
and the twisted conjugation, both yielding unitarity. Several 
applications are scrutinized, involving the 
inherent darkness underlying ELKO 
construction and galaxies rotation curves, 
the non-relativistic potential regime, and 
the M{\o}ller-like scattering involving 
ELKO.
\end{abstract}
\maketitle
\section{Introduction}
Quantum fields constructed upon mass dimension one quantum spinor fields in QFT  are prime candidates to describe dark matter, since they are, by construction, neutral fermions under gauge interactions, therefore implementing darkness \cite{Ahluwalia:2022ttu,Ahluwalia:2019etz,Ahluwalia:2018hfm}. Neutrality under gauge fields is naturally implemented when one employs eigenspinors of the charge conjugation operator. Additionally demanding  that the right- and left-handed components of these spinor fields have dual helicity, the so-called  eigenspinors of the
charge conjugation operator with dual helicity (ELKO) set in. Despite its inherent darkness, eventual couplings remain and might lead to peculiar experimental signatures of ELKO in high-energy running experiments  \cite{Lee:2015sqj,Alves:2014kta,Alves:2014qua,Alves:2017joy,Duarte:2020svn,Dias:2010aa,Agarwal:2014oaa,BuenoRogerio:2017zxf,Duarte:2017svd,Rogerio:2016mxi,HoffdaSilva:2019eao}.
ELKO resides in non-standard Wigner spinor  classes, when the unitary irreducible representations of the Poincar\'e group are extended to  encode  discrete symmetries \cite{Ahluwalia:2020jkw, Ahluwalia:2004ab,Ahluwalia:2004sz,Ahluwalia:2008xi,Ahluwalia:2009rh}. 
 Field-theoretical developments and applications of  mass dimension one quantum fields were presented in Refs. \cite{Bernardini:2012sc,Fabbri:2019vut,Fabbri:2017xyk,Fabbri:2014foa,daRocha:2011yr,Lee:2018ull,Lee:2015tcc,Lee:2015jpa,Fabbri:2012yg,Fabbri:2011pbj,Fabbri:2011mi,Fabbri:2010qv,Fabbri:2010ws,Fabbri:2009aj,daRocha:2009gb,Nieto:2022rdo,Cavalcanti:2014uta,deBrito:2019hih,Nieto:2019qyw,Rogerio:2022tsl,Dale:2022jub,HoffdaSilva:2022ixq,daRocha:2008we,Fabbri:2020nmx,daRocha:2007pz,Fabbri:2020elt,BuenoRogerio:2019ict,HoffdaSilva:2016ffx,BuenoRogerio:2016seo,Lee:2014opa,Lee:2012td,Vaz:2017fac,Nikitin:2014fga}. In addition, fermionic aspects of AdS/CFT correspondence, emulating mass dimension one quantum fields,  have been introduced in Refs.   \cite{Meert:2018qzk,deBrito:2016qzl,Bonora:2015ppa,Bonora:2014dfa,Lopes:2018cvu,Yanes:2018krn}, whereas the pivotal role of ELKO in the membrane paradigm was investigated in Refs. \cite{Dantas:2015mfi,Zhou:2017bbj,Liu:2011nb,MoazzenSorkhi:2020fqp,Jardim:2014xla,Zhou:2018oib,Maluf:2019ujv}. Other aspects of ELKO in gravity can be found in Ref. \cite{Vignolo:2021frk}. The  
Hawking radiation of ELKO across the horizon of several black hole solutions was studied in Refs. \cite{daRocha:2014dla,daRocha:2016bil,Cavalcanti:2015nna}. Cosmological aspects of ELKO were addressed in Refs. \cite{Pereira:2017efk,Kouwn:2012eat,Pereira:2017bvq,Pereira:2016eez,Basak:2014qea,Basak:2012sn,Sadjadi:2012xyd}.

Some limitations of using a QFT for ELKO, when Maxwell gauge fields are employed, can be circumvented when  electrodynamics with a Poldolsky sector is taken into account in the ELKO paradigm, together with a generalized spinor dual. As Podolsky himself  asserted, 
the only way that Maxwell electrodynamics can be generalized consists of allowing the Lagrangian to contain terms involving higher-order derivatives of the electromagnetic field strength  \cite{Podolsky:1942zz}. Introducing it naturally imposes a cutoff procedure that precludes undesirable effects in QFT due to higher frequencies. Also, in the QFT setup with a Podolsky sector, there is an  additional gauge freedom arising from higher-order field equations, allowing  to remove any singularity that is intrinsic to the standard approach,  avoiding divergences like the one occurring due to the vacuum polarization current and the 
electron self-energy as well \cite{Bertin:2009gs}. The main goal of this work is to investigate ELKO scattering processes, including the ELKO-photon and ELKO-proton interaction, in standard QFTs with Maxwell fields and, subsequently,  also involving a  Podolsky sector, yielding generalized QFTs. Renormalization and radiative corrections are also implemented in the ELKO setup, being unitarity also constructed.  Several possible applications of this approach are also pointed out and explored, involving the inherent darkness underlying ELKO construction and galaxies rotation curves. This work is organized as follows: 
Sec. \ref{sec1} is devoted to revisit ELKO setup as 
 eigenspinors of the
charge conjugation operator with dual helicity. ELKO quantum fields are also introduced, using a generalized spinor dual and the associated creation and annihilation operators. The spin sums are constructed through an odd-parity operator. The ELKO Hamiltonian structure is also presented. Cross-sections and scattering processes involving ELKO are still reviewed in this section. 
In Sec. \ref{sec2}, the spin sums and the Feynman rules for ELKO-photon interaction, in both Podolsky and Maxwell QFTs, are studied and discussed. Some intricacies involving the polarization tensor and bubble diagrams involving ELKO and anti-ELKO are presented, considering the Dirac spinor dual. Unitarity aspects are scrutinized.
In Sec. \ref{sec3} the Podolsky sector in a QFT for ELKO  is introduced, being the associated propagator explored, together with radiative corrections. 
Renormalization features are also discussed.
Sec. \ref{sec4} is dedicated to reanalyzing the content in Secs. \ref{sec2} and \ref{sec3} using the ELKO generalized spinor dual, instead of the Dirac spinor dual. The unitarity is analyzed and the comparison between different spinor field conjugations is implemented, using the new Hermitian-like conjugation prescription \cite{Ahluwalia:2022ttu}, the so-called  twisted  conjugation. Within this setup, the polarization tensor regarding ELKO is computed and divergences are controlled by renormalization procedures. The 3-point function is also calculated, ELKO self-interaction is included and decay rates are studied, also employing the Podolsky propagator. The associated beta functions are also computed. Sec. \ref{sec6} discusses  Podolsky QFT and unitarity aspects involving ELKO scattering processes, regarding an effectively
combined propagator. In 
Sec. \ref{sec7}, the vertexes, the gauge propagator, and the external ELKO, are  studied in the low-energy approach, to obtain the non-relativistic potential.  A Podolsky gauge field sector and the ELKO fermionic sector with the generalized dual are employed, yielding the associated scattering amplitude to be derived. The potential is shown to be a Yukawa-type one. Relevant aspects are then discussed, as the dark particle system  does not interact at large distances.  
Sec. \ref{sec8} scrutinizes  the corrected M{\o}ller scattering with the generalized ELKO spinor dual as well as the twisted conjugation. 
Sec. \ref{sec9} is devoted to furnishing useful tools for evaluating the relic density and the freeze-out temperature, with the scattering amplitude for the
ELKO pair annihilation being computed, and in Sec. \ref{snew} the ELKO-proton scattering is addressed, with the analysis of an effective vertex. 
In Sec. \ref{sec10}, a linear term arising from the self-energy corrections for the model composed of ELKO spinors with the Dirac dual  is employed to study a solution exhibiting a galaxy flat rotation curve. It is implemented by analyzing the ELKO fermionic response associated with the exotic theory constructed with the Dirac spinor dual instead of the generalized  ELKO dual, with the addition of a Podolskyan propagator. Some interesting conclusions are made by comparing our results and structures with some analogous content for topological insulator description in condensed
matter. Finally in 
Sec. \ref{sec11} the concluding remarks 
are presented as well as additional discussion and perspectives.  The metric signature $(+,-,-,-)$ is used throughout.

\section{ELKO underlying framework and ramifications}
\label{sec1}
To approach QFT, in general one takes a spinor $\psi(p^\mu)$ in the $\left(\frac12,0\right)\oplus\left(0,\frac12\right)$ irreducible representation of the Lorentz group to constitute spin-1/2 quantum fields, which in the Weyl representation of the gamma matrices reads
\begin{equation}
		\psi(p^\mu) = \left( \begin{array}{c}
		\upphi_+(p^\mu)\\
		\upphi_-(p^\mu)
		\end{array}
		\right).
\end{equation}
The right-  and left-handed Weyl spinors, respectively carrying the $\left(\frac12,0\right)$ and  $\left(0,\frac12\right)$ Lorentz group irreducible representations, are eigenspinors of the helicity operator, 
\begin{equation}
(\s\cdot\widehat \p) \upphi_\pm(k^\mu) = \pm \upphi_\pm(k^\mu),
\end{equation}
where, in spherical coordinates, 
\begin{align}
\upphi_+(k^\mu) & = \sqrt{m}  \left(
									\begin{array}{c}
									\cos\left(\frac{\theta}{2}\right)e^{- i \phi/2}\\
									\sin\left(\frac{\theta}{2}\right)e^{i \phi/2}
								\end{array}
	\right),\qquad\qquad
\upphi_-(k^\mu) = \sqrt{m}   \left(		\begin{array}{c}
									- \sin\left(\frac{\theta}{2}\right)e^{- i \phi/2}\\
									 \cos\left(\frac{\theta}{2}\right)e^{i \phi/2}
											\end{array}
									\right),
\end{align}
for $k^\mu=\lim_{p\to0}p^\mu$ \cite{Ahluwalia:2022ttu}.
To introduce mass dimension one fermions \cite{Ahluwalia:2004sz,Ahluwalia:2004ab,Ahluwalia:2016rwl,Ahluwalia:2019etz}, one must better investigate the spinor dual, defined as $
\overline{\psi}(p^\mu) \,{=}\, \psi^\dagger(p^\mu)
\upeta.$  
The spin-1/2 matrix  $\upeta$ is determined from the constraint that bilinears constructed from spinors are  covariant. 
The matrix $\upeta$ commutes with the generators of rotation and anticommutes with the boost generators, supporting the local gauge transformations of the Standard Model. For the Dirac spinor field, one usually takes  $\upeta = \gamma^0$. 
For ELKO, the definition is distinct, as dark matter fields have to carry irreducible representations of the Lorentz group, with also well-defined properties under the parity ($\mathfrak{P}$), charge conjugation ($\mathfrak{C}$), and time reversal ($\mathfrak{T}$) operators. ELKO do not support (local) gauge transformations of the Standard Model. ELKO are defined as  eigenspinors of charge conjugation operator  \cite{Ahluwalia:2004sz,Ahluwalia:2004ab,Ahluwalia:2016rwl,Ahluwalia:2019etz}, 
\begin{equation}
\uplambda_\upalpha(p^\mu) {=} \left(\begin{array}{c}
\upzeta_\uplambda \Uptheta\left[ \upphi_\upalpha(p^\mu)\right]^\ast \\
\upphi_\upalpha(p^\mu)
\end{array}
\right),
\label{eq:ELKO}
\end{equation}
where $\Uptheta=\scriptsize{\begin{pmatrix}-1&0\\0&1\end{pmatrix}}$ is the Wigner time reversal operator satisfying $\Uptheta\left(\sigma_i\right)\Uptheta^{-1}=-\sigma_i^*$, for each Pauli matrix $\sigma_i$.
Here $\upalpha$ (and throughout this paper upgreek indexes also) denotes the helicity index. Their right- and left-hand components present opposite helicity, 
\begin{equation}
 \s\cdot\widehat{\p} \left[\Uptheta \upphi_\pm^\ast(p^\mu)\right]  = \mp
  \left[\Uptheta \upphi_\pm^\ast(p^\mu)\right]  .
 \end{equation}
As posed in Ref. \cite{Ahluwalia:2019etz}, the charge conjugation operator reads 
\begin{equation}
	\mathfrak{C} = \left(\begin{array}{cc}
					\0 & i\Uptheta \\
					-i \Uptheta & \0
					\end{array}
					\right) K, 
	\end{equation}
where the $K$ operator implements complex conjugation. Choosing $\upzeta_\uplambda = +i$ [$-i$] yields a self-conjugate $\uplambda^S(p^\mu)$ [anti-self-conjugate $\uplambda^A(p^\mu)$] ELKO, 
\begin{equation}
\mathfrak{C} \uplambda^{S}_\pm(p^\mu) =  +  \uplambda^{S}_\pm(p^\mu),\qquad
\mathfrak{C} \uplambda^{A}_\pm(p^\mu) =   - \uplambda^{A}_\pm(p^\mu).
\end{equation}
Hence the spinors  at rest, $\uplambda_\upalpha(k^\mu)$, read  \cite{Ahluwalia:2019etz}
\begin{align}
 &\uplambda^S_\pm(k^\mu)  =  \left(
					\begin{array}{c}
					+i \Uptheta\left[\upphi_\pm(k^\mu)\right]^\ast\\
								\upphi_\pm(k^\mu)
					\end{array}
					\right),  \qquad\qquad\uplambda^A_\pm(k^\mu)  =  \left(
					\begin{array}{c}
					- i \Uptheta\left[\upphi_\mp(k^\mu)\right]^\ast\\
								\upphi_-(k^\mu)
					\end{array}
					\right),
				\end{align}
For arbitrary momentum, ELKO are obtained from the rest spinors by the action of the
 boost transformation
as
 $
\uplambda^{S/A}_{\pm}(p^\mu) = D(L(p)) \; \uplambda^{S/A}_{\pm}(k^\mu).$ 
One defines spin-1/2 quantum fields with ELKO as its expansion coefficients \cite{Ahluwalia:2022ttu,Ahluwalia:2019etz,Ahluwalia:2020miz,Ahluwalia:2008xi}, 
\begin{equation}
\mathring{\mathfrak{f}}(x) =\int
\frac{d^3 p}{(2\pi)^3}
\frac{1}{\sqrt{2 m E(\p)}}
\sum_\upalpha
\left[
a_\upalpha(\p)
\uplambda^S_\upalpha(\p) e^{-i p\cdot x}
+
b^\dagger_\upalpha(\p)
\uplambda^A_\upalpha(\p) e^{i p\cdot x}
\right].                   \label{eq113}
\end{equation}
 The spinor field $\mathfrak{f}(x)$ has {mass dimension} 
one and fermionic statistics. Therefore ELKO fields cannot partake Standard Model doublets, therefore constituting first-principle candidates to describe dark matter \cite{Ahluwalia:2022ttu,Ahluwalia:2020jkw,Ahluwalia:2019etz,Ahluwalia:2018hfm}.
To construct the Feynman--Dyson propagator associated with ELKO, a  twisted spinor dual is necessary, since 
the norm of ELKO, with respect to the Dirac dual, equals zero. In fact, $
\overline{\uplambda}^{S/A}_\upalpha(p^\mu) {\uplambda}^{S/A}_\upalpha(p^\mu)
=0.$ 
Therefore a new spinor dual was introduced in Ref. \cite{Ahluwalia:2019etz}, 
\begin{align}
& \tilde{\uplambda}_{\pm}^{S/A}(p^\mu) 
=  \mp i \left[ \uplambda^S_{\mp} (p^\mu) \right]^\dagger
 \gamma_0, \label{eq114}
\end{align}
yielding the following orthonormality  relations
\begin{align}
\tilde\uplambda^S_\upalpha(p^\mu) \uplambda^S_{\upalpha^\prime}
(p^\mu) = 2 m \delta_{\upalpha\upalpha^ \prime}=-\tilde\uplambda^A_\upalpha(p^\mu) \uplambda^A_{\upalpha^\prime}
(p^\mu),\qquad\qquad\quad \tilde\uplambda^S_\upalpha(p^\mu) \uplambda^A_{\upalpha^\prime}(p^\mu) = 0 =
\tilde\uplambda^A_\upalpha(p^\mu) \uplambda^S_{\upalpha^\prime}(p^\mu),
\end{align}
with the associated spin sums
\begin{align}
& \sum_\upalpha \uplambda^S_\upalpha(p^\mu) \tilde\uplambda^S_\upalpha(p^\mu) = m \left[\I+\mathfrak{G}(p^\mu)
\right], \qquad\quad
\sum_\upalpha \uplambda^A_\upalpha(p^\mu) \tilde\uplambda^A_\upalpha(p^\mu) = -m \left[\I-\mathfrak{G}(p^\mu)
\right],\label{ss1}
\end{align}
taking into account the odd-parity operator
\begin{equation}
\mathfrak{G}(p^\mu) = i
\left(
\begin{array}{cccc}
0 & 0 &  0 &- e^{-i\phi}\\
0 & 0 & e^{i\phi} & 0\\
0 & - e^{-i\phi} &0 &0 \\
 e^{i\phi} & 0 & 0 &0
\end{array}
\right).\label{op}
\end{equation}
The obtained spin sums lead to the completeness relation
\begin{align}
\frac{1}{2m}\sum_{\upalpha}\bigg[\uplambda^S_\upalpha(p^\mu)
\tilde{\uplambda}^S_\upalpha(p^\mu) -
\uplambda^A_\upalpha(p^\mu)
\tilde{\uplambda}^A_\upalpha(p^\mu)\bigg] = \I.
\end{align}
The matrix $\mathfrak{G}(p^\mu)$ yields a preferred direction, therefore violating  locality~\cite{Ahluwalia:2010zn}, which can be circumvented by redefining the ELKO dual \cite{Ahluwalia:2022ttu,Ahluwalia:2019etz,Ahluwalia:2018hfm},
\begin{align}
&\tilde{\uplambda}^S_\upalpha(p^\mu) 
\mapsto \gdualn{\uplambda}^S_\upalpha(p^\mu) {=}\tilde{\uplambda}^S_\upalpha(p^\mu) \mathfrak{A}, 
&\tilde{\uplambda}^A_\upalpha(p^\mu) \mapsto \gdualn{\uplambda}^A_\upalpha(p^\mu) = \tilde{\uplambda}^A_\upalpha(p^\mu) \mathfrak{B},\label{eq116}
\end{align}
where $\mathfrak{A}$ and $\mathfrak{B}$ are operators such that the self-conjugate ELKO are   eigenspinors of $\mathfrak{A}$, 
whereas anti-self-conjugate ELKO are   eigenspinors of the $\mathfrak{B}$ operator, both corresponding to unity eigenvalues.
Also, the conditions
\begin{equation}
\tilde{\uplambda}^S_\upalpha(p^\mu)\mathfrak{A} \uplambda^A_{\upalpha^\prime}(p^\mu)=0= \tilde{\uplambda}^A_\upalpha(p^\mu)\mathfrak{B} \uplambda^S_{\upalpha^\prime}(p^\mu)
\end{equation}
hold as well. With these two operators,  the orthonormality relations read
\begin{align}
& \gdualn\uplambda^S_\upalpha(p^\mu) \uplambda^S_{\upalpha^\prime}(p^\mu)
 =  2 m \delta_{\upalpha\upalpha^\prime}= - \gdualn\uplambda^A_\upalpha(p^\mu) \uplambda^A_{\upalpha^\prime}(p^\mu), &  \gdualn\uplambda^S_\upalpha(p^\mu) \uplambda^A_{\upalpha^\prime}(p^\mu) = 0 =
 \gdualn\uplambda^A_\upalpha(p^\mu) \uplambda^S_{\upalpha^\prime}(p^\mu),
 \label{eq117}
\end{align}
whereas the spin sums for both the self- and anti-self-conjugate ELKO  read
\beq
\sum_{\upalpha} \uplambda^S_\upalpha(p^\mu) \gdualn{\uplambda}^S_\upalpha(p^\mu) &=& 
m \big[\I + \mathfrak{G}(p^\mu) \big] \mathfrak{A},
\qquad\quad
\sum_{\upalpha} \uplambda^A_\upalpha(p^\mu) \gdualn{\uplambda}^A_\upalpha(p^\mu)  =
- m \big[\I - \mathfrak{G}(p^\mu)\big] \mathfrak{B}. \label{ss3}
\eeq
Ref. \cite{Ahluwalia:2019etz} showed that the $\mathfrak{A}$ and $\mathfrak{B}$ operators can be derived from $\mathfrak{G}(p^\mu)$ by
\begin{equation}
\mathfrak{A} = 2\lim_{\uptau\to1} \left[\I + \uptau\mathfrak{G}(p^\mu)\right]^{-1},\quad\qquad
\mathfrak{B} = 2\lim_{\uptau\to1} \left[\I - \uptau\mathfrak{G}(p^\mu)\right]^{-1},
\end{equation}
where $\uptau$ is a real parameter.
This leads to spin sums
\begin{align}
& \sum_\upalpha\uplambda^S_\upalpha(p^\mu) \gdualn{\uplambda}^S_\upalpha(p^\mu) =  2 m \I = - \sum_\upalpha \uplambda^A_\upalpha(p^\mu) \gdualn{\uplambda}^A_\upalpha(p^\mu),
\end{align}
with completeness relation $
\sum_{\upalpha}\left[ \uplambda^S_\upalpha(p^\mu) \gdualn{\uplambda}^S_\upalpha(p^\mu) - \uplambda^A_\upalpha(p^\mu) \gdualn{\uplambda}^A_\upalpha(p^\mu)\right]  = 4m\I.$

ELKO are well known not to satisfy Dirac equations, but the following coupled system of first-order field equations \cite{Dvoeglazov:1995eg,Dvoeglazov:1995kn,Ahluwalia:2019etz}, 
\begin{align}
&\gamma_\mu p^\mu \uplambda^S_\pm(p^\mu) = \pm im \uplambda^S_\mp(p^\mu),\qquad\quad
 \gamma_\mu p^\mu \uplambda^A_\pm(p^\mu) =\mp  im \uplambda^A_\mp(p^\mu),\label{coupled}
\end{align} satisfying the Klein--Gordon equations, 
\begin{equation}
\left(g_{\mu\nu}p^\mu p^\nu \I - m^2 \I\right)\uplambda^{S/A}_{\upalpha}(p^\mu) =0.
\end{equation}
To prospect the statistics satisfied by the ELKO field $\mathfrak{f}(x)$  in Eq. (\ref{eq113}), its adjoint is naturally defined as
\begin{equation}
\mathring{\gdualn{\mathfrak{f}}}(x) =
\int
\frac{d^3 p}{(2\pi)^3}
\frac{1}{\sqrt{2 m E(\p)}}
\sum_\upalpha\left[
a^\dagger_\upalpha(\p) \gdualn{\uplambda}^S_\upalpha(\p) e^{i p\cdot x}
+ b_\upalpha(\p) \gdualn{\uplambda}^A_\upalpha(\p) e^{- i p\cdot x}
\right].\label{dual}
\end{equation}
To preclude any causal paradox, Ref. \cite{Ahluwalia:2015vea} showed that the ELKO and its adjoint have a vanishing  anticommutator, 
$
\{\mathring{\gdualn{\mathfrak{f}}}(x) , \mathring{\mathfrak{f}}(x^\prime)\}=0$, 
with  fermionic statistics to the creation and annihilation operators, yielding a consistent local QFT. 
Ref. \cite{Ahluwalia:2019etz} showed that the Feynman--Dyson propagator reads
\begin{align}
S_{{\scalebox{0.6}{$\textsc{FD}$}}}(x^\prime-x) & = 
-\frac{i}{2} \langle~\vert
\mathfrak{T}[\;\mathfrak{f}(x^\prime)\mathfrak{f}(x)\;]\vert~\rangle =
\int\frac{d^4p}{(2\pi)^4}\frac{\I}{\Box + m^2 + i\epsilon}e^{i p\cdot(x - x^\prime)},
\end{align} with mass dimension one and free field Lagrangian density given by
\begin{equation}
\mathcal{L}_{{\scalebox{0.6}{$\textsc{free}$}}}(x) = \frac{1}{2}\Big(g_{\mu\nu}\partial^\mu {\gdualn{\mathfrak{f}}} (x)
\partial^\nu\mathfrak{f}(x)
 - m^2 {\gdualn{\mathfrak{f}}}(x) \mathfrak{f}(x)\Big). \label{eq:Lagrangian}
\end{equation} 
The $\mathfrak{f}(x)$ field contributes to the zero-point field energy as
\begin{align}
&H_{{\scalebox{0.6}{$\textsc{free}$}}}^\mathfrak{f} = -4\int d^3 x\int \frac{d^3p}{{(2\pi)^3}}\;\frac{1}{2} E(\p).
\end{align}
Using the parity operator $\mathfrak{P}= m^{-1} \gamma_\mu p^\mu$ \cite{Speranca:2013hqa}, and the time-reversal operator $\mathfrak{T}= i \gamma^{5}\mathfrak{C}$, the properties of Elko under $
 (\mathfrak{C}\mathfrak{P}\mathfrak{T})^2 = \I$ and $\left\{\mathfrak{C},\mathfrak{P}\right\}=0,$
indicate that ELKO quantum fields belong to the non-standard  Wigner spinor classes \cite{Ahluwalia:2004ab,Ahluwalia:2004sz}. 
Lounesto's spinor field classification unites all spinor fields in 4-dimensional spacetime into six disjoint classes, according to the values attained by the bilinear covariants computed from the respective spinor field. This classification is based upon the U(1) gauge 
symmetry of the first-order equations of motion that rule
spinor fields in each spinor class. Lounesto's classification devises mass dimension one spinors as a class of singular spinors \cite{daRocha:2005ti}. A more general classification has been proposed in Ref. \cite{Fabbri:2017lvu} encompassing spinor multiplets as realizations of (non-Abelian) gauge fields.
A reciprocal classification, generalizing Lounesto's idea, was implemented in Refs. \cite{Cavalcanti:2014wia,Ablamowicz:2014rpa}.
Mass dimension one quantum fields were investigated under the prism of non-standard Wigner classes in Refs.  
\cite{HoffdaSilva:2012uke,Cavalcanti:2020obq,BuenoRogerio:2019yqb,BuenoRogerio:2019kgd,HoffdaSilva:2019ykt,Beghetto:2018mkp,BuenoRogerio:2020bdm,daRocha:2007sd,Rogerio:2020cqg,daRocha:2013qhu}, also paving the way for other generalized spinor field classifications \cite{Arcodia:2019flm,DaRocha:2020oju,CoronadoVillalobos:2020yvr,HoffdaSilva:2017waf}, including the spinor classification encompassing higher order gauge groups \cite{Bonora:2017oyb}. Flipping phenomena between spinor field classes were analyzed
in Ref. \cite{HoffdaSilva:2019xvd}.

In QFT, a transition probability can be computed when Hermitian conjugation is employed, to certify a positive-definite probability. Ref. \cite{Lee:2015sqj} showed that the optical theorem is violated at 1-loop,  implementing quantum corrections to field theory, and  coming from the non-Hermitian character of the ELKO mass dimension one fermion. Therefore, Ref. \cite{Ahluwalia:2022ttu} introduced the twisted conjugation $\ddag$, required to be an involution,  to compute transition probabilities and observables. The ELKO dual (\ref{eq114}) can be recast as
\begin{equation}
\tilde{\uplambda}^{S/A}_{\upalpha}(\p)=\left[\uplambda^{S/A}_{\upalpha}(\p)\right]^{\ddag}\gamma_{0}=-i\upalpha\left[\uplambda^{S/A}_{-\upalpha}(\p)\right]^{\dag}\gamma_0.\label{eq119}
\end{equation}
Besides, Ref. \cite{Ahluwalia:2022ttu} showed that 
\begin{align}
\left[\uplambda^{S/A}_{\upalpha}(\p)^{\ddag}\uplambda^{S/A}_{\upalpha'}(\p')\right]^{\ddag}
=i\upalpha\left[\uplambda^{S/A}_{\upalpha'}(\p')\right]^{\ddag}\left[\uplambda^{S/A}_{-\upalpha}(\p)^{\dag}\right]^{\ddag}.
\label{eq121}
\end{align}
Employing the twisted conjugation $\ddag$, the expansion of the ELKO quantum field and its dual can be rewritten as 
\begin{subequations}
\begin{align}
&\mathfrak{f}(x)= \int\frac{d^{3}p}{{(2\pi)^3}\sqrt{2 m E(\p)}}\sum_{\upalpha}\left[a_{\upalpha}(\p)\uplambda^{S}_{\upalpha}(\p)e^{-ip\cdot x}+b^{\ddag}_{\upalpha}(\p)\uplambda^{A}_{\upalpha}(\p)e^{-ip\cdot x}\right],\\
&\gdualn{\mathfrak{f}}(x)= \int\frac{d^{3}p}{{(2\pi)^3}\sqrt{2 m E(\p)}}\sum_{\upalpha}\left[a^{\ddag}_{\upalpha}(\p)\gdualn{\uplambda}^{S}_{\upalpha}(\p)e^{ip\cdot x}+b_{\upalpha}(\p)\gdualn{\uplambda}^{A}_{\upalpha}(\p)e^{-ip\cdot x}\right].
\end{align}
\end{subequations}
The annihilation and creation operators satisfy the canonical anticommutators
\begin{equation}
\left\{a_{\upalpha}(\p),a^{\ddag}_{\upalpha'}(\p')\right\}=\left\{b_{\upalpha}(\p),b^{\ddag}_{\upalpha'}(\p')\right\}=(2\pi)^{3}\delta_{\upalpha\upalpha'}\delta^{3}(\p-\p'),
\end{equation}
with $a^{\ddag\ddag}_{\upalpha}(\p)\equiv a_{\upalpha}(\p)$ and $b^{\ddag\ddag}_{\upalpha}(\p)\equiv b_{\upalpha}(\p)$. 
Although the Lagrangian is non-Hermitian, it satisfies $\mathcal{L}_{{\scalebox{0.6}{$\textsc{free}$}}}^{\ddag}(x)=\mathcal{L}_{{\scalebox{0.6}{$\textsc{free}$}}}(x)$. For mass dimension one fermions, the usual demand of Hermiticity should be replaced by the demand of invariance under the twisted conjugation $\ddag$.

One can therefore compute observables employing the $\ddag$ operator. Ref. \cite{Ahluwalia:2022ttu} considers scattering  processes regarding ELKO for  $\mathcal{M}_{ba}$ denoting the scattering amplitude. Denoting by  $\mho_{ba}$ a unimodular phase function, the 2-body cross-section for the $12\rightarrow34$ scattering process reads 
\begin{align}
\sigma\big\vert_{12\rightarrow34}
=&\frac{(2\pi)^{4}\delta^{4}(p_{3}+p_{4}-p_{1}-p_{2})}{16\sqrt{(p_{1}\cdot p_{2})^{2}\!-\!m^{2}_{1}m^{2}_{2}}}\int\!\!\frac{d^{3}p_{3}}{(2\pi)^{3}E_{3}}\int\!\!\frac{d^{3}p_{4}}{(2\pi)^{3}E_{4}}
\left[\mho_{{\scalebox{0.7}{$(34)(12)$}}}\mathcal{M}^{\ddag}_{{\scalebox{0.7}{$(34)(12)$}}}\mathcal{M}_{{\scalebox{0.7}{$(34)(12)$}}}\right].
\end{align}
Interactions set in by regarding renormalizable interaction potentials, denoting by $\phi(x)$ a scalar field,  
\begin{eqnarray}
&& V_{\phi}(t)=\frac{g_{\phi}}{2}\int d^{3}x[\gdualn{\mathfrak{f}}(x)\mathfrak{f}(x)\phi^{2}(x)],\qquad\quad  V_{\mathfrak{f}}(t)=\frac{g_{\mathfrak{f}}}{2}\int d^{3}x[\gdualn{\mathfrak{f}}(x)\mathfrak{f}(x)]^{2}.
\end{eqnarray}
The transition probability regarding ELKO as external states (the capital letter indexes below labeling self-conjugate and anti-self-conjugate ELKO),
\begin{equation}
K_{IJ}\equiv \left[\gdualn{\uplambda}^{I}_{\alpha}(\p)\uplambda^{J}_{\alpha'}(\p')\right]\left[\gdualn{\uplambda}^{J}_{\alpha'}(\p')\uplambda^{I}_{\alpha}(\p)\right].
\end{equation}
When $I=J$ [$I\neq J$] two external fermions or antifermions [an external fermion-antifermion pair] manifest, whereas
$K_{AA}=K_{SS}$ are non-negative quantities and $K_{AS}=K_{SA}$ are non-positive ones. 
Therefore $\mathcal{M}^{\ddag}_{ba}\mathcal{M}_{ba}$ is positive-definite [negative-definite] for  even [odd] number of external antifermions, yielding $\mho_{ba}=(-1)^{\mathring{n}_{ba}},$ 
for $\mathring{n}_{ba}$ being the total number of incoming and outgoing antifermions. 


It is worth to mention that for the Dirac dual the following expression holds,
\bea \sum_\upalpha \uplambda_\upalpha^{S/A}(p) \bar \uplambda_\upalpha^{S/A}(p)=  \slashed{p}.        \eea
There is a relation between ELKO under parity, as\footnote{The Levi--Civita symbol regarding helicity indexes reads $\epsilon_{+-} = +1 = -\epsilon_{-+}$.}
\bea \uplambda^{S/A}_\upalpha(p^\mu)=i\epsilon_{\upalpha \upalpha^\prime}\uplambda_{\upalpha^\prime}^{A/S}(-p^\mu),       \eea
 implying that  
\bea \sum_\upalpha \uplambda_\upalpha^{S/A}(p^\mu) \bar \uplambda_\upalpha^{S/A}(p^\mu)=\sum_\upalpha \uplambda_\upalpha^{A/S}(-p^\mu) \bar \uplambda_\upalpha^{A/S}(-p^\mu),    \eea
which is also valid for the spin sum of the generalized ELKO dual (\ref{eq116}).

Taking the free Lagrangian  \eqref{eq:Lagrangian} into account and regarding  $\mathcal{L}_{{\scalebox{0.6}{$\textsc{free}$}}}$ in (\ref{eq:Lagrangian}), the conjugate momentum to $\mathfrak{f}(x)$ reads \cite{Ahluwalia:2022ttu,Ahluwalia:2019etz}
\begin{equation}
\mathfrak{p}(x) =
\frac{\partial \mathcal{L}_{{\scalebox{0.6}{$\textsc{free}$}}}(x)}
{\partial       \dot{\mathfrak{f}}   (x)   } = \frac{1}{2} \frac{\partial}{\partial t}
\gdualn{\mathfrak{f}}(x),
\qquad  \qquad \gdualn{\mathfrak{p}}(x) =
\frac{\partial \mathcal{L}_{{\scalebox{0.6}{$\textsc{free}$}}}(x)}
{\partial       \dot{\gdualn{\mathfrak{f}}}  (x)   } = \frac{1}{2} \frac{\partial}{\partial t}
{\mathfrak{f}}(x).
\end{equation}
The locality of ELKO is  regulated by 
\begin{align}
\left\{
\mathfrak{f}(t,x),\mathfrak{p}(t,x^\prime)\right\} &=
i \delta^3(x-x^\prime), \qquad \left\{
\gdualn{\mathfrak{f}}(t,x),\gdualn{\mathfrak{p}}(t,x^\prime)\right\} = 
i \delta^3(x-x^\prime), \nonumber \\
0&=\left\{
\mathfrak{f}(t,x),\mathfrak\mathfrak{f}(t,x^\prime)\right\} = 
\left\{
\mathfrak{p}(t,\x),\mathfrak{p}(t,\x^\prime)\right\} = 0.
\end{align} 
The on-shell Hamiltonian density can be obtained \cite{Ahluwalia:2004ab}. Here, $\mathfrak{f}$ and $\gdualn{\mathfrak{f}}$ are considered to be independent variables as well as in the Dirac particle case, 
\bea {\cal{H}}= \dot{\gdualn{\mathfrak{f}}}\mathfrak{p}+\gdualn{\mathfrak{p}}\dot{\mathfrak{f}} -{\cal{L}_{{\scalebox{0.6}{$\textsc{free}$}}}}=2\dot{\gdualn{\mathfrak{f}}}\dot{{\mathfrak{f}}}.\eea
The Hamiltonian operators reads
\bea H=2 \int \frac{d^3p}{(2\pi)^32mE(\p)}E^2(\p)\sum_{\upalpha,\beta}\Big(\bar \uplambda^S_\upalpha(p^\mu) \uplambda^S_\beta(p^\mu) a^\dagger_\upalpha a_\beta+ \bar \uplambda^S_\upalpha(p^\mu) \uplambda^S_\beta(p^\mu) b_\upalpha b_\beta^\dagger     \Big)\nn\\ =2\int \frac{d^3p}{(2\pi)^3}E(\p)\sum_{\upalpha \beta}i \epsilon_{{\upalpha^\prime}\upalpha}\left(a^\dagger_\upalpha a_{\upalpha^\prime}+b^\dagger_{\upalpha^\prime}b_\upalpha\right),         \eea being Hermitian, with eigenstates 
\beq A_\upalpha^\dagger|0\rangle =\left(a_\upalpha^\dagger+i\epsilon_{\upalpha \upalpha^\prime}a_{\upalpha^\prime}^\dagger\right)|0\rangle,\qquad  B_\upalpha^\dagger|0\rangle=\left(b_\upalpha^\dagger+i\epsilon_{\upalpha \upalpha^\prime}b_{\upalpha^\prime}^\dagger\right)|0\rangle.\eeq The second eigenstate has negative energy. In order to circumvent it, we consider $\{a^\dagger_\upalpha,a_{\upalpha^\prime}\}=\{b^\dagger_\upalpha,b_{\upalpha^\prime}\}=\delta_{\upalpha \upalpha^\prime}$ and, for the antiparticle, we flip the interpretation to consider $b_\upalpha^\dagger$ as the anihilation operator and $b_\upalpha$ as the creation one. Therefore, the antiparticle eigenstate becomes $B_\upalpha|0\rangle=(b_\upalpha+i\epsilon_{\upalpha\upalpha^\prime}b_{\upalpha^\prime})|0\rangle$. In this case, both have positive energy and also positive norm. Since the eigenstates are of the form $2E(p)$, one can normalize the ELKO field (\ref{eq113}) as $\frac{1}{\sqrt{2}} \mathfrak{f}$ to $E(\p)$ be the Hamiltonian eigenvalue. \\
\indent For the case of the generalized ELKO dual (\ref{eq116}), the Hamiltonian reads
\bea H=2\int \frac{d^3p}{(2\pi)^3}E(\p)\sum_{\upalpha}\left(a^\dagger_\upalpha a_\upalpha+b^\dagger_\upalpha b_\upalpha\right),        \eea
which is analogous to the mass dimension three-halves fermionic case.

\section{Feynman rules for ELKO-photon interaction and unitarity aspects}
\label{sec2}
This section is devoted to present the Feynman rules regarding the various types of ELKO and to investigate associated aspects regarding unitarity of a QFT describing also ELKO\footnote{Hereone Eq. (\ref{eq117}) will be expressed as \begin{align}
& \gdualn\uplambda^S_\upalpha(p^\mu) \uplambda^S_{\upalpha^\prime}(p^\mu)
 =  m \delta_{\upalpha\upalpha^\prime}= - \gdualn\uplambda^A_\upalpha(p^\mu) \uplambda^A_{\upalpha^\prime}(p^\mu), &  \gdualn\uplambda^S_\upalpha(p^\mu) \uplambda^A_{\upalpha^\prime}(p^\mu) = 0 =
 \gdualn\uplambda^A_\upalpha(p^\mu) \uplambda^S_{\upalpha^\prime}(p^\mu),
 \label{altern}
\end{align}
which is equivalent to considering each type of ELKO divided to a $\sqrt{2}$ factor.}. Employing the Dirac dual, the Feynman rules related to self-conjugate and anti-self-conjugate ELKO spinors are respectively given by
\beq
 \frac{\bar \uplambda^S_\upalpha(p^\mu)}{\sqrt{m}}
    =  \vcenter{\hbox{\includegraphics[width=1.4cm,height=0.85cm]{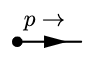}}}
    \qquad\qquad\qquad\frac{\uplambda^S_\upalpha(p^\mu)}{\sqrt{m}}
    =  \vcenter{\hbox{\includegraphics[width=1.4cm,height=0.85cm]{ext1.png}}}\nn\\
    \frac{\bar \uplambda^A_\upalpha(p^\mu)}{\sqrt{m}}
    = \vcenter{\hbox{\includegraphics[width=1.35cm,height=0.9cm]{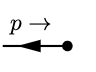}}}\qquad\qquad\qquad
      \frac{\uplambda^A_\upalpha(p^\mu)}{\sqrt{m}}
    =  \vcenter{\hbox{\includegraphics[width=1.35cm,height=0.9cm]{ext3}}} \eeq
       whereas the polarizations account for
       \begin{eqnarray} 
     \epsilon_{\mu }^{*\upalpha}(p)
    =  \vcenter{\hbox{\includegraphics[width=1.6cm,height=1cm]{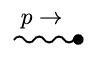}}}
       \qquad\qquad\qquad
     \epsilon_{\mu }^{\upalpha}(p)
    = \vcenter{\hbox{\includegraphics[width=1.6cm,height=1cm]{pol1.png}}} \end{eqnarray}
       The ELKO field propagator can be obtained using the methodology of Ref. \cite{Ahluwalia:2004ab} for the Dirac dual case   
       \bea \vcenter{\hbox{\includegraphics[width=3.2cm,height=1.15cm]{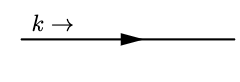}}}=\frac{i\slashed{k}}{m\Big( k^2-m^2+i\epsilon \Big)}\eea
The photonic sector is represented, up to gauge fixing terms,  by    
\begin{equation}
\vcenter{\hbox{\includegraphics[width=2.6cm,height=1.4cm]{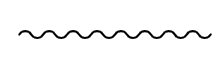}}}=i\mathfrak{G}_{\mu \nu}(k)=-\frac{ig_{\mu \nu}}{k^2+i\epsilon}+A\frac{ig_{\mu \nu}}{k^2-M^2+i\epsilon}.
\end{equation}
 The cases with $A=1$ (Podolskyan) and $A=0$ (Maxwellian) will be considered.

The vertex for ELKO-photon interaction, governed by the Lagrangian 
\beq{\cal{L}}_I=\frac{i}{2}\bar{\mathfrak{f}}(x) F_{\mu \nu}(x)\sigma^{\mu \nu}\mathfrak{f}(x)={\cal{L}}_I^\dagger,\eeq
reads
\begin{equation}\vcenter{\hbox{\includegraphics[width=3.1cm,height=2.6cm]{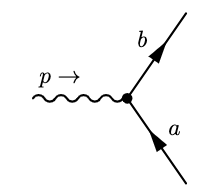}}}=ig p_\mu\sigma^{\mu \nu}_{{ba}}
       \end{equation}
with $\sigma_{\mu \nu}\equiv \frac{1}{2}\left[\gamma_\mu,\gamma_\nu \right]$. The $a$ and $b$ latin letters denote spinor indexes. Considering the Dirac dual, the polarization tensor reads
\begin{equation}\begin{aligned}
 i\mathcal{\pi}_{\mu \nu}(p)=-\frac{g^2}{m^2}  \int \frac{d^4k}{(2\pi)^4}\frac{Tr\big( p_\omega\sigma^{\omega \mu}(\slashed{p}-\slashed{k})p_\alpha \sigma^{\alpha \nu}\slashed{k})}{\big( k^2-m^2+i\epsilon \big)\big( (p-k)^2-m^2+i\epsilon \big)}=
\end{aligned}
\vcenter{\hbox{\includegraphics[width=3.8cm,height=2.45cm]{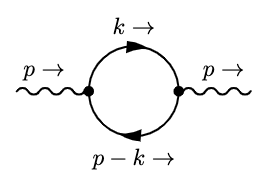}}}
\qquad\qquad
\end{equation}
  
The imaginary part of a massive photon that decays into an ELKO and an anti-ELKO which combine again to form a photon, \beq\epsilon^\mu2\Im\big[i\mathcal{\pi}_{\mu \nu}(p) \big ] \epsilon^{*\nu},\eeq is given by 
\beq
\!\!\!\!\!\!\!\!\!\!\!\!-\frac{g^2}{m^2}\int \frac{d^4k}{(2\pi)^2}\theta(k_0)\theta(p_0-k_0)\epsilon_\mu \epsilon^{*}_\nu Tr\big( p_{\alpha}\sigma^{\alpha \mu}(\slashed{p}-\slashed{k})p_{\beta}\sigma^{\beta\nu}\slashed{k} \big)\delta\left((p-k)^2-m^2\right)      \delta \left( k^2-m^2\right).  
\eeq
To verify the optical theorem, we must find the formal expression to the associated decay rate
\beq
   \!\!\!\!\!\!\!\!\!\!\! \lim_{M\to 0}\! \sum_{\upalpha,\upalpha^\prime}\!\int\!\!\frac{d^4k}{(2\pi)^2} \theta(k_0)\theta(p_0-k_0) \delta\Big(k^2\!-\!m^2 \Big)\delta\Big((p\!-\!k)^2\!-\!M^2 \Big) 
    \vcenter{\hbox{\includegraphics[width=4.2cm,height=2.18cm]{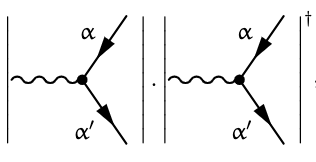}}}
       \eeq
       which equals $2M_{{\scalebox{0.6}{$\textsc{photon}$}}}\Gamma$. The photon mass $M_{{\scalebox{0.6}{$\textsc{photon}$}}}$ equals zero in the case here studied. However, the formal expression must be fulfilled in order to ensure unitarity. The spin sum  over the squared amplitudes yields
\begin{multline}
\vcenter{\hbox{\includegraphics[width=4.2cm,height=3cm]{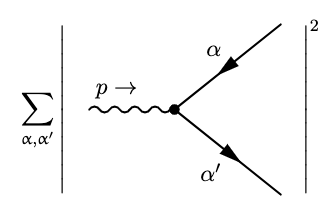}}}=-\left(\frac{1}{\sqrt{m}}\right)^4\!\!g^2\sum_{\upalpha,\upalpha^\prime} \epsilon^\mu \epsilon^{*\nu}\bar \uplambda_{\upalpha^\prime}^S(k)p_\omega \sigma^{\omega \mu}\uplambda_\upalpha^A(p-k)\bar \uplambda_\upalpha^A(p-k) p_\omega \sigma^{\omega \nu}\uplambda_{\upalpha^\prime}^S(k)
       \\=-\frac{g^2}{m^2}\epsilon^\mu \epsilon^{\nu *} Tr\big( p_{\zeta}\sigma^{\zeta \mu}(\slashed{p}-\slashed{k})p_{\beta}\sigma^{\beta\nu}\slashed{k} \big),\end{multline}
\!where the spin sums, the Feynman rules, and the expression $\sigma_{\mu \nu}^\dagger=-\gamma_0\sigma_{\mu \nu}\gamma_0$ have been considered. Therefore, the optical theorem is verified. Denoting $\theta_{\mu \nu}=g_{\mu \nu}-\frac{p_\mu p_\nu}{p^2}$, one can write the imaginary part of the polarization tensor as \beq 2
\Im\big[i\mathcal{\pi}_{\mu \nu}(p) \big ]\!=\!\theta_{\mu \nu}\Pi(p^2),\eeq in which
\beq \!\!\!\!\!\!\!\!\!\!\!\!\!\! {3\Pi(p^2)}\!&\!=\!&\!2g^{\mu \nu}\Im\big[i\mathcal{\pi}_{\mu \nu}(p) \big ]\nonumber\\\!\!\!\!\!\!\!\!&\!=\!
&\!-\frac{g^2}{m^2}\int\! \frac{d^4k}{(2\pi)^2}\theta(k_0)\theta(p_0-k_0)g_{\mu \nu} Tr\big( p_{\zeta}\sigma^{\zeta  \mu}(\slashed{p}\!-\!\slashed{k})p_{\beta}\sigma^{{\beta}\nu}\slashed{k} \big)\delta\big((p\!-\!k)^2\!-\!m^2\big)      \delta \big(k^2\!-\!m^2\big).\eeq
Together with gamma matrices identities and the relations implied by the delta functions as well, it yields 
\beq   3\Pi(p^2)=-\frac{g^2}{m^2}\int \frac{d^4k}{(2\pi)^2}\theta(k_0)\theta(p_0-k_0) 2p^2\big(p^2+2m^2     \big)\delta\big((p-k)^2-m^2\big)      \delta \big( k^2-m^2\big).          \eeq
This object has the same sign as the one from QED. This is an important check to ensure unitarity, which will be useful when analyzing the possibility of a bosonic sector composed of the so-called Podolsky model. Therefore, as a consistency check, considering an external photon with polarization $\upalpha$  and the fact that these transverse excitations obey $\epsilon^{\mu\upalpha}\epsilon_\mu^{*\upalpha}=-1$, the expected positivity of the decay rate becomes manifest, for the case of a massive photon. Otherwise, for a massless one, it vanishes\footnote{This feature is going to be useful in the next sections regarding a harmless introduction of the Podolsky field.}.
 Considering still the Dirac dual, the self-energy has the following form, 
 \begin{multline}
     i\Sigma(p)=\frac{g^2}{m}\int \frac{d^4k}{(2\pi)^4}\frac{(p-k)_\beta\sigma^{\beta \gamma} \slashed{k}(p-k)_\zeta\sigma^{\zeta}_{\;\, \gamma}\big(-M^2\big)^N}{\big( k^2-m^2+i\epsilon \big)\big( (p-k)^2+i\epsilon \big)\big( (p-k)^2-M^2+i\epsilon \big)^N }\\  \\  \vcenter{\hbox{\includegraphics[width=4.7cm,height=2.2cm]{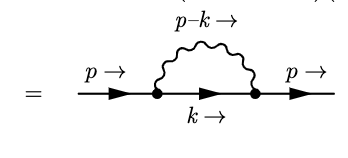}}}\quad \quad \quad \quad \quad \quad \quad \quad \quad \quad \quad \quad \quad \quad \quad \quad \quad \quad \quad \quad \quad \quad \quad \quad \quad \quad \quad \quad
    \end{multline}
\!\!with $N=0$ for Maxwell and $N=1$ for the Podolsky QFT.  Now, for $N=0$, the imaginary part for amplitude of a process associated to an incoming particle that emits a photon, reabsorbs it and then continues to propagate, reads
${\bar{\uplambda}}^S(p)2\Im\big[\Sigma(p)  \big ] \uplambda^S(p)$, or equivalently,
\beq \!\!\!\!\!\!\!\!\! \frac{g^2}{m}\int \frac{d^4k}{(2\pi)^2}\theta(k_0)\theta(p_0-k_0)
\bar{\uplambda}^S(p)(p-k)_\omega \sigma^{\omega \mu} \slashed{k}(p-k)_{\beta}\sigma^{\beta}_{\;\, \mu} \uplambda^S(p)\delta\big((p-k)^2\big)      \delta \big(k^2-m^2\big).   \eeq
To verify the optical theorem, one must evaluate the total decay rate,
\begin{multline}
    \lim_{M\to 0} \sum_{r,s}\!\int\!\frac{d^4k}{(2\pi)^2} \theta(k_0)\theta(p_0-k_0) \delta\left(k^2\!-\!m^2 \right)\delta\left((p\!-\!k)^2\!-\!M^2 \right)
       =\vcenter{\hbox{\includegraphics[width=4.2cm,height=2.18cm]{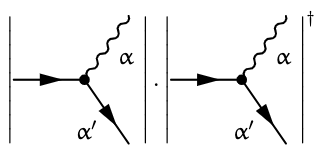}}}=2m \Gamma.\nn
       \end{multline}
The explicit spin sum over the squared amplitude is given by

\begin{multline}
\vcenter{\hbox{\includegraphics[width=4.2cm,height=2.9cm]{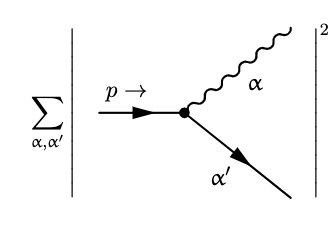}}}=-\frac{g^2}{\sqrt{m}^2}\sum_{\upalpha,\upalpha^\prime} \bar \uplambda^S(p)(p-k)_\omega \sigma^{\omega \mu}\uplambda_{\upalpha^\prime}^S(k)\bar \uplambda_{\upalpha^\prime}^S(k) (p-k)_\omega \sigma^{\omega \nu}\uplambda^S(p)\epsilon_\mu^\upalpha \epsilon_\nu^{\upalpha*}\\
       =\frac{g^2}{m}\bar \uplambda^S(p)(p-k)_\omega \sigma^{\omega \mu}\slashed{k} (p-k)_\zeta \sigma^{\zeta }_{\; \mu}\uplambda^S(p).
       \end{multline}
      \!Here the identities $\sum_\upalpha\epsilon_\mu^{\upalpha}\epsilon_\nu^{\upalpha*}=-g_{\mu \nu}+\frac{p_\mu p_\nu}{p^2}$ have been used. Therefore, the unitarity is verified. The same conclusion can be obtained for the amplitude of an anti-conjugate ELKO  that emits a photon, absorbs it and then continues to propagate. The spin sum relevant to compute the decay rate 

\beq
\vcenter{\hbox{\includegraphics[width=4.5cm,height=3.2cm]{squared2.png}}}
       \eeq
       
       represents the term
       \beq&&-\frac{1}{\sqrt{m}^2}g^2\sum_{{\upalpha,\upalpha^\prime}} \bar \uplambda^A(-p)(p-k)_\zeta \sigma^{\zeta \mu}\uplambda_{\upalpha^\prime}^A(-k)\bar\uplambda_{\upalpha^\prime}^A(-k) (p-k)_\omega \sigma^{\omega \nu}\uplambda^A(-p)\epsilon_\mu^\upalpha \epsilon_\nu^{\upalpha*}\nn\\
        &&\qquad\qquad=\frac{g^2}{m}\bar \uplambda^A(p)(p-k)_\zeta \sigma^{\zeta \mu}\slashed{k} (p-k)_\omega \sigma^{\omega }_{\;\, \mu} \uplambda^A(p).\eeq
       We have used the property relating a self-conjugate spinor evaluated in $p$ with an anti-self-conjugate spinor at $-p$.
The imaginary part of the amplitude reads
\beq && \bar \uplambda^A(-p)2\Im\big[\Sigma(p)  \big ] \uplambda^A(-p),\eeq or, equivalently,  
\beq\frac{g^2}{m}\int \frac{d^4k}{(2\pi)^2}\theta(k_0)\theta(p_0-k_0)\bar \uplambda^A(-p)(p-k)_\omega \sigma^{\omega \mu} \slashed{k}(p-k)_{\beta}\sigma^{\beta}_{\;\, \mu} \uplambda^A(-p)\delta\big((p-k)^2\big)      \delta \big( k^2-m^2\big).        \eeq
Therefore, this process complies with unitarity.  It is worth mentioning that the operator expansion coefficients of the quantum field are not the energy eigenstates\footnote{In fact, they are a linear combination of these operators.}, in opposition to the generalized  ELKO dual case, see \ref{sec2}. Despite this fact, investigating this exotic Dirac dual possibility is still relevant, due to its associated unitary QFT and the existence of a solution with potential application in the description of dark matter phenomenology \cite{Pereira:2016emd}. Interestingly, these microscopic and cosmological aspects are intimately related.

\section{Renormalization properties}
\label{sec3}
The use of the Dirac dual leads to a propagator with the high energy behavior as $1/p$ and not $1/p^2$, as in the case with the generalized ELKO dual (\ref{eq116}) \cite{Ahluwalia:2004ab}. It means that the investigation of the renormalization properties is relevant. First of all, one can write
\bea    ig^{\mu \nu}\mathcal{\pi}_{\mu \nu}(p^2)\equiv \mathcal{\pi}(p^2)=12(p^2)^2\frac{g^2}{m^2}  \int \frac{d^4k}{(2\pi)^4}dx\frac{x(1-x)}{\big( k^2-\Delta(p^2)+i\epsilon \big)^2},           \eea
with $\Delta(p^2)=m^2-p^2x(1-x)$.  Since $\mathcal{\pi}_{\mu \nu}(p^2)=\theta_{\mu \nu}\frac{\mathcal{\pi}(p^2)}{3}$, we conclude that to eliminate the logarithmic divergence, a Podolsky sector  must be included in the bare Lagrangian. Denoting the electromagnetic field strength by $F_{\mu\nu} = \partial_{\mu}A_{\nu}-\partial_\nu A_\mu$, the Podolsky model can be written as \footnote{Its well known form can be obtained by simply Gaussian integrating the $B_\mu(x)$ field. }
\bea
    {\cal{L}}_{{\scalebox{0.6}{$\textsc{pod}$}}}=-\frac{1}{4}F_{\mu \nu}F^{\mu \nu}-\frac{a^2}{2}B_\mu B^\mu+a^2\partial_\mu B_\nu F^{\mu \nu},\label{pod1}
\eea
with $a^2=\frac{1}{M^2}$,  for $M$ denoting the Podolsky photon mass, and $B_\mu$ denoting the electromagnetic potential in the Podolsky sector. To decouple the fields, one can redefine $A_\mu \mapsto A_\mu+a^2 B_\mu$. Absorbing the factor $a^2$, $B_\mu \mapsto a^2B_\mu$,  yields
\bea
    {\cal{L}}_{{\scalebox{0.6}{$\textsc{pod}$}}}=-\frac{1}{4}F_{\mu \nu}F^{\mu \nu}-\frac{1}{2}B^\mu \big(M^2g_{\mu \nu}+\Box \theta_{\mu \nu}  \big) B^\nu.\label{pod2}
\eea
The self-energy has the structure given by \footnote{As presented in the next sections, the extra massive field associated to the Podolsky  sector  does not violate the unitarity constraints previously investigated. This is due to the fact that the model is an interacting one leading to a Merlin mode behaviour for $B_\mu(x)$.}
\beq i\Sigma(p)=\frac{g^2}{m}\int dx \frac{d^4k}{(2\pi)^4}\frac{[p(1-x)-k]_\alpha\sigma^{\alpha \gamma} \big(\slashed{k}+\slashed{p}x\big)[p(1-x)-k]_\beta\sigma^{\beta}_{\;\, \gamma}}{\big( k^2-\tilde \Delta(p^2)+i\epsilon \big)^2 }-\Sigma \big(p,\tilde \Delta(p^2,M^2)\big),\eeq
with \beq\tilde \Delta(p)=(1-x)(m^2-p^2x), \qquad\qquad \tilde \Delta(p,M)\equiv (1-x)(m^2-p^2x)-xM^2. \eeq The subtracted terms account for a possible use of the Podolsky propagator.
 The divergent parts read 
\beq
i\Sigma_{{\scalebox{0.6}{$\textsc{div}$}}}(p)=\frac{g^2}{m}\int \frac{d^4k\ dx}{(2\pi)^4}\frac{[\frac{9}{2}(x-1)k^2\slashed{p}+3p^2\slashed{p}x(1-x)^2]}{\big( k^2-\tilde \Delta(p^2)+i\epsilon \big)^2 }-\Sigma_{{\scalebox{0.6}{$\textsc{div}$}}} \big(p,\tilde \Delta(p^2,M^2)\big).\eeq 

\indent If a conventional photon propagator is used, there are quadratic divergences associated with a $\slashed{p}$ term and a logarithmic one associated with $p^2\slashed{p}$. In the case of a Podolsky propagator, just the linear term in external momenta is logarithmic divergent and the remaining ones are finite. This result differs from the bare Lagrangian and demands an extra renormalization condition. On the other hand,  it seems to be the only kind of divergent term, even in higher-order loops. It means that just one extra condition is necessary.  Therefore, the fermionic 2-point part is renormalizable,  since it needs a finite set of renormalization conditions. Therefore, considering the case of a Podolsky propagator yields the following divergent piece for the linear contribution
\beq \Sigma_{{\scalebox{0.6}{$\textsc{div}$}}}^{{\scalebox{0.6}{$\textsc{linear}$}}}(p)&=&-\slashed{p}\frac{(2-D)D}{4}\frac{g^2M^2}{m}\int dx \frac{9(x-1)x}{2}\frac{\mu^{-\epsilon}}{16\pi^2} \left[-\frac{2}{\epsilon}+\gamma_E\right]\nn\\&=&-\slashed{p}\frac{3M^2g^2}{m}\frac{\mu^{-\epsilon}}{32\pi^2} \left[-\frac{2}{\epsilon}+\gamma_E\right]. \eeq
It is equivalent to an effective action with an extra term $i\bar f A\slashed{p}f$. It varies with the energy as
\bea  \mu \frac{dA}{d\mu}= -\frac{3M^2g^2}{16\pi^2m}.  \eea This term may lead to interesting solutions associated with the observed flat rotation curve of the galaxies. Even for the case of the Maxwell propagator, the cubic term in $p$ can be discarded at low energies in a cosmological scenario. Therefore, the linear term is the focus of our attention.

The next important object to investigated is related to the radiative corrections for the interaction vertex. The 1-loop contribution is given by
\beq 
i\Gamma_\mu&=& g^3\int \frac{1}{(2\pi)^4}d^4k\; dx dy dz\;\delta \big( x+y+z-1\big)\frac{k_\gamma^1\sigma^{\gamma \alpha}\slashed{k}_3\big(\sigma^{{\beta}\mu}p_\beta\big) k^4_\zeta \sigma^{\zeta}_{\; \alpha}\slashed{k}^9  }{[k^2-\Delta'(p)]^3}          -i\Gamma_\mu(\Delta')\vert_M,\eeq
where $\Delta'(p)\equiv m^2(1-z)^2-xyp^2$ and $\Delta'(p,M)\equiv \Delta'(p)+zM^2$, 
which can be illustrated by the process

\begin{equation}
\vcenter{\hbox{\includegraphics[width=6cm,height=2.7cm]{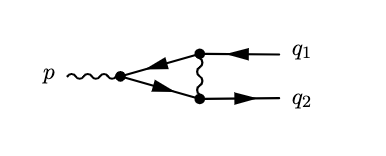}}}
\end{equation}
       
 whereas the momentum variables are defined as
\beq
k^1_\gamma &\equiv& k_\gamma-yp_\gamma+(z-1)q^1_\gamma, \quad\qquad\quad k^3_\gamma\equiv k_\gamma-yp_\gamma+zq^1_\gamma, \nn\\k^4_\gamma &\equiv& k_\gamma-(y+1)p_\gamma+zq^1_\gamma, \quad\qquad\quad   k^9_\gamma \equiv k_\gamma-yp_\gamma+(z-1)q^1_\gamma.      \eeq 
Again, the subtracted term represents the effect of using a Podolsky propagator associated to mixture of massless and massive photons with mass $M$. 
Considering the Podolsky propagator, the only divergence is of logarithmic kind and proportional to the integrand of
\bea \sum_a\int \frac{d^4k}{(2\pi)^4}\frac{\ldots\slashed{k}\big(\sigma^{{\upalpha^\prime}\mu }p_{\upalpha^\prime}\big)\slashed{k}\ldots}{[k^2-\Delta'_a]^3}.   \eea
This sum is taken over the contributions from the gauge field propagator parts with massless and massive poles, associated to $a=1,2$, respectively. 
If the integration symmetry is considered, $\slashed{k}\big(\sigma^{{\beta}\mu }p_{\beta}\big)\slashed{k}=\frac{k^2}{4}\gamma^\mu \big(\sigma^{{\beta}\mu }p_{\beta}\big)\gamma_\mu=0 $, since  $\gamma^\nu \big(\sigma^{{\upalpha^\prime}\mu }p_{\upalpha^\prime}\big)\gamma_\nu=0 $. Therefore, no counterterm is necessary for this object.

Regarding the 4-point functions, if the Podolsky model is employed, there is no divergence, and our 1-loop analysis is completed. On the other hand, using the Maxwell propagator leads to divergent terms coming from the 4-point function. These are different from the bare Lagrangian ones. New divergent terms may arise at higher loops implying in a non-renormalizable theory since there are infinite divergent anomalous terms each one demanding its renormalization condition to be eliminated. 

\section{ELKO generalized dual, unitarity and comparison between different spinor field conjugations}
\label{sec4}
The propagator associated with the generalized ELKO dual \eqref{eq116} reads
\begin{equation} \vcenter{\hbox{\includegraphics[width=3.5cm,height=1.2cm]{propagator.png}}}=\frac{i}{k^2-m^2+i\epsilon}\end{equation}
Regarding the generalized  ELKO dual (\ref{eq116}), the spin sum associated to the decay rate, calculated via the Hermitian conjugation $\dagger$, related to the polarization tensor, reads

\begin{multline}
\vcenter{\hbox{\includegraphics[width=4.5cm,height=3.2cm]{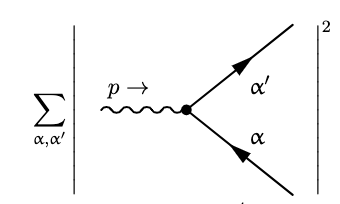}}}\!=\!-\left(\!\frac{1}{\sqrt{m}}\!\right)^4\!g^2\ \sum_{\upalpha,\upalpha^\prime} \epsilon^\mu \epsilon^{*\nu}\gdualn{\uplambda}_{\upalpha^\prime}^S(k)p_\zeta \sigma^{\zeta \mu}\uplambda_\upalpha^A(p-k)i\epsilon_{\upalpha\upbeta}\gdualn{\uplambda}_\upbeta^A(p-k) p_\omega \sigma^{\omega \nu}i\epsilon_{\upalpha^\prime \updelta}\uplambda_{\updelta}^S(k)
       \\=-\frac{g^2}{m^2}\epsilon_\mu \epsilon_\nu^* Tr\big( p_{\alpha}\sigma^{\alpha \mu}(\slashed{p}-\slashed{k})p_{\beta}\sigma^{{\beta}\nu}\slashed{k} \big).\end{multline}
       Computing the imaginary part of the associated polarization tensor yields 
\beq \!\!\!\!\!\!\!\!\!\epsilon^\mu2\Im\big[i\mathcal{\pi}_{\mu \nu}(p) \big ] \epsilon^{*\nu}\!=\!
g^2\!\int\!\! \frac{d^4k}{(2\pi)^2}\theta(k_0)\theta(p_0\!-\!k_0)\epsilon_\mu \epsilon_\nu Tr\big( p_{\alpha}\sigma^{\alpha \mu}p_{\beta}\sigma^{{\beta}\nu}\big)\delta\big((p\!-\!k)^2\!-\!m^2\big)      \delta \big(k^2\!-\!m^2\big).  \eeq
Therefore, under the Hermitian conjugation $\dagger$, the theory with the generalized  ELKO dual is not compatible with the optical theorem and  violates unitarity. A generalized  conjugation prescription $\ddagger$, as introduced in Sec. \ref{sec1}, must be then regarded. It is also important to mention that, considering this generalized prescription $\ddagger$, the radiative corrections of the model are in agreement with the optical theorem and lead to a unitary theory.

Renormalization aspects of the model composed of the generalized ELKO dual can be therefore implemented. 
One can now regard the $\ddagger$ twisted conjugation acting on the spinor indices as defined in Eq. \eqref{eq119}. 
The Lagrangian structure is invariant under the $\ddagger$ operator, 
\bea {\cal{L}}=\frac{g}{2}i\df \sigma_{\mu \nu} F^{\mu \nu}{\mathfrak{f}}={\cal{L}}^\ddagger.\label{ll1} \eea

The explicit divergent parts of the associated polarization tensor reads 
\bea i\mathcal{\pi}_{\mu \nu}(p)=g^2  \int \frac{d^4k}{(2\pi)^4}\frac{Tr\big( p_\omega\sigma^{\omega \mu}p_\alpha \sigma^{\alpha \nu}\big)}{\big( k^2-m^2+i\epsilon \big)\big( (p-k)^2-m^2+i\epsilon \big)}=4ip^2g^2\theta_{\mu \nu}\frac{\mu^{-\epsilon}}{16\pi^2}\left(  -\frac{2}{\epsilon}+\gamma_E\right). \eea
 In the last equality just the divergent and constant terms, that are eliminated in a $\bar{MS}$ renormalization scheme, have been regarded. Differently from the Dirac dual case, there is no necessity of using a Podolsky  bosonic sector. On the other hand, its use does not generate any inconsistency and also improves the renormalization properties. 
 First of all, the standard QFT with  Maxwell electrodynamics will be analyzed. The divergent part of the polarization tensor can be absorbed by a renormalization of the photonic field. Denoting the square of the photon renormalization constant as $Z_p=1+\delta_1$ and noticing that $\mathcal{\pi}_{\mu \nu}^R=\mathcal{\pi}_{\mu \nu}+\delta_1p^2\theta_{\mu \nu}$, it reads $\delta_1=-4g^2\frac{\mu^{-\epsilon}}{16\pi^2}\big(  -\frac{2}{\epsilon}+\gamma_E\big) $. 
The divergent piece of the self-energy can be expressed as
\beq i\Sigma(p)&=&g^2\int \frac{d^4k}{(2\pi)^4}\frac{(p-k)_\alpha\sigma^{\alpha \gamma} (p-k)_\beta\sigma^{\beta}_{\;\, \gamma}}{\big( k^2-m^2+i\epsilon \big)\big( (p-k)^2+i\epsilon \big) }\nn\\
&=&-g^2\int \frac{d^4k}{(2\pi)^4}\frac{3}{\big( k^2-m^2+i\epsilon \big) }=3im^2g^2\frac{\mu^{-\epsilon}}{16\pi^2}\left(-\frac{2}{\epsilon}+\gamma_E \right).\eeq
 Therefore, since\footnote{The $\delta_2$ is associated to the ELKO spinor wave function renormalization and $\delta_m$ with the renormalization constant regarding the mass, $Z_m$.} \beq\Sigma^R(p)=\Sigma(p)+p^2\delta_2-m^2(\delta_2+\delta_m),\eeq then $\delta_2=0$ and \beq\delta_m m^2=3m^2g^2\frac{\mu^{-\epsilon}}{16\pi^2}\left(-\frac{2}{\epsilon}+\gamma_E \right).\eeq
 Besides, the 3-point function reads
\begin{multline} i\Gamma_\mu= ig^3\int \frac{d^4k }{(2\pi)^4} \frac{(k-q_1)_\gamma\sigma^{\gamma \alpha}\big(i\sigma^{{\beta}\mu}p_\beta\big) (k-q_1)_\zeta \sigma^{\zeta}_{\;\, \alpha}  }{[k^2-m^2][(k-q_1)^2][(k+p)^2-m^2]}=ig^3\int \frac{d^4k }{(2\pi)^4} \frac{i\sigma^{\beta\mu}p_\beta  }{[k^2-m^2][(k+p)^2-m^2]}\quad \quad \quad \quad \quad \quad \quad \\
=g^3\big(i\sigma^{{\beta}\mu}p_\beta\big) \frac{\mu^{-\epsilon}}{16\pi^2}\left(-\frac{2}{\epsilon}+\gamma_E\right),\quad \quad \quad \quad \quad \quad \quad \quad \quad \quad \quad \quad \quad \quad \quad \end{multline}
where 
\beq\Gamma_\mu^R=\Gamma_\mu+g\left(\delta_g+\frac12\delta_1\right)\gamma_{\beta\mu}p^\beta.\eeq
Hence one can write
\beq\delta_g+\frac12\delta_1=-g^2 \frac{\mu^{-\epsilon}}{16\pi^2}\left(-\frac{2}{\epsilon}+\gamma_E\right),\eeq with
$\delta_g$ being associated to the renormalization of the coupling constant. Therefore, we have $\delta_g=g^2 \frac{\mu^{-\epsilon}}{16\pi^2}\big(-\frac{2}{\epsilon}+\gamma_E\big)$. Since the bare coupling can be written as\footnote{$Z_g=1+\delta_g$} $g^0=Z_g g^R$, it yields
\bea    \mu \frac{\partial g_R}{\partial \mu}=-\frac{g_R^3}{8\pi^2}.       \eea
Therefore, one can digress about asymptotic freedom for this model. As mentioned, the model complies with unitarity and the optical theorem, if one use the twisted $\ddagger$ conjugation, as well as for the Dirac dual case.  The imaginary part of the polarization tensor can be written as $2\Im\big[i\mathcal{\pi}_{\mu \nu}(p) \big ]=\theta_{\mu \nu}\Pi(p^2)$. The scalar $\Pi(p^2)$ is evaluated as
\beq \Pi(p^2)&=&\frac{2}{3}g^{\mu \nu}\Im\big[i\mathcal{\pi}_{\mu \nu}(p) \big ]\nn\\&=
&-4g^2\int \frac{d^4k}{(2\pi)^2}\theta(k_0)\theta(p_0-k_0)p^2\delta\big((p-k)^2-m^2\big)      \delta \big( k^2-m^2\big).\eeq As aforementioned, the model is unitary. Hence, associating this result with the formal expression for the decay rate of the photon\footnote{For a massless photon it is zero, for a massive one, it is indeed positive.}, and considering the polarization identities $\epsilon^{*\upalpha\mu}\epsilon_\mu^\upalpha=-1$ yields a positive $\Gamma$. For the case of ELKO self-interaction \cite{Ahluwalia:2004ab, Ahluwalia:2004sz,Ahluwalia:2022ttu},
\begin{equation}
\left({\gdualn{\mathfrak{f}}}(x) \mathfrak{f}(x)\right)^2,
\end{equation}  
 an analogous decay associated with ELKO and anti-ELKO would be negative and a sign must be included to ensure a probabilistic interpretation. The reason for the absence of an extra sign for the decay rate is due to the fact that Hermitian conjugating the vertex $\sigma^{\mu \nu}p_\mu$ leads also to a minus sign. Therefore, when calculating decay rates, one must consider for the probability density
 \beq
 P=\sum_b \mathcal{M}_{ab}\mathcal{M}_{ba}^\ddagger (-1)^N(-1)^V,\eeq with $N$ being the number of antifermions in $\mathcal{M}_{ab}$ and $V$ denoting the number of vertexes in the  amplitude. The fact that the previously mentioned formal decay is positive is important,  if one wants to use the Podolsky propagator without any unitarity or causality violation.  If one considers a box with four external fermions, there are divergent terms that are different from the bare Lagrangian. Their number may increase if one considers higher loops. The only way to turn the system into a renormalizable one is to consider the Podolsky model for the bosonic sector. In terms of divergences, the use of the Podolsky propagator\footnote{It is given by $-i\left(\frac{1}{p^2}-\frac{1}{p^2-M^2} \right)=-i\frac{(-M^2)}{p^2(p^2-M^2)}$.} leads to a finite $\Gamma_\mu$, finite external four (or more) legs diagram and a divergent self-energy as
\begin{multline} i\Sigma_{{\scalebox{0.6}{$\textsc{div}$}}}^{{\scalebox{0.6}{$\textsc{pod}$}}}(p)=-3(-M^2)g^2\int \frac{d^4k}{(2\pi)^4}\frac{1}{\big( k^2-m^2\big)\big( (p-k)^2-M^2\big)}=-3M^2g^2\frac{i\mu^{-\epsilon}}{16\pi^2}\left(-\frac{2}{\epsilon}+\gamma_E\right).  \end{multline}
In this case, $\delta_2=0$ and \beq
m^2(\delta_2+\delta_m)=-3g^2 M^2 \frac{\mu^{-\epsilon}}{16\pi^2}\left(-\frac{2}{\epsilon}+\gamma_E \right).\eeq Therefore, \beq \delta_m=-3\frac{g^2M^2}{m^2}\frac{\mu^{-\epsilon}}{16\pi^2}\left(-\frac{2}{\epsilon}+\gamma_E \right).\eeq Since the vertex correction is not divergent, the identity  $\delta_g+\frac{\delta_1}{2}+\delta_2=0$ must hold\footnote{With $\delta_1$ coming from the one loop polarization tensor renormalization. It has the same value for both Maxwell and Podolsky models.}, leading to $\delta_g=2g^2\frac{\mu^{-\epsilon}}{16\pi^2}\big(-\frac{2}{\epsilon}+\gamma_E \big)$. Therefore, it yields a factor that is the half of the one regarding the previously investigated Maxwellian beta function \footnote{For the case of the Podolsky model, the quadratic term in derivatives, proportional to $a^2\equiv \frac{1}{M^2}$, also contributes to the counterterms associated to the photon 2-point correlation function. Due to the structure of the 1-loop polarization tensor, one must have $a^2\left( \delta a^2+\delta_1\right)=0$ with $a^2=Z_{a^2}a^2_R$. }
\bea    \mu \frac{\partial g_R}{\partial \mu}=-\frac{g_R^3}{4\pi^2}.       \eea

\section{Podolsky theory and unitarity}
\label{sec6}
This section is devoted to discussing the physical implications that the Podolsky theory which, as we have seen, can be derived when one lowers the order of the derivatives, similarly to a Maxwell theory coupled to a massive Proca model, with a different  sign. Therefore, the free theory necessarily leads to unitarity violation \cite{Donoghue:2021eto}. On the other hand, if the model is interacting, the imaginary part of the polarization tensor evaluated at the Podolsky mass does not necessarily vanish. If this object has the right sign, it leads to a kind of resonance called Merlin mode \cite{Donoghue:2021eto}.  The normal massless resonance propagates  forward in time, with positive energy, whereas the higher-mass
pole propagates backward,  distinguishing the Merlin mode  from the ghost, which has a minus sign in the numerator appearing in  the propagator. 
Faddeev--Popov ghosts present a negative sign in the numerator, however, they carry the usual imaginary unit, in the denominator. In the case of the Merlin mode, a change of sign occurs in the denominator as well,  yielding the associated propagator to represent the time-reversed version
of the standard resonance propagator. It does not violate unitarity, just microcausality in a small time scale of the form $t\sim 1/|\gamma|$, with $|\gamma|$ being the modulus of the imaginary part evaluated at the propagator pole mass. If it has the opposite sign, this contribution grows indefinitely leading to ill behavior. Therefore, since both theories with ELKO and Dirac duals have a positive definite formal expression for the photon decay, the addition of the Podolsky content can be consistently carried out for these models.

Unitarity is not violated in this case since the lines of a decaying particle are never cut, and do not generate external asymptotic particles. Therefore, we do not consider the external massive photons in our Feynman rules.  
 Up to gauge fixing and longitudinal sectors, the inverse propagator related to the massive sector, associated to $B_\mu(x)$ defined in Sec. \ref{sec3}, evaluated near the renormalized mass, reads  \bea     {\cal{P}}_{\mu \nu}^{-1}=-ig_{\mu \nu}(p^2-M^2)-i\mathcal{\pi}_{\mu \nu},\                                                  \eea
yielding, up to longitudinal non-physical terms, 
\bea {\cal{P}}_{\mu \nu}=\frac{ig_{\mu \nu}}{\big(p^2-M^2_r-i|\gamma|\big)},                   \eea
with $\mathcal{\pi}_{\mu \nu}\equiv \theta_{\mu \nu}\Pi$, reflecting the fact that the imaginary part of the trace of the polarization tensor is negative for the QFTs here studied, as well as  QED, if they are evaluated for a time-like external momentum as, for example, the case of a massive propagator evaluated near the pole. Then, this massive bosonic excitation becomes a Merlin mode\footnote{$|\gamma|$ is greater than zero if $M_r>2m$.} \cite{Donoghue:2021eto}, 
contributing to improve the renormalizability properties of the system without violating the unitarity. Since both $A_\mu$ and $B_\mu$ appearing in the Podolsky model (\ref{pod1}, \ref{pod2}) are coupled to the ELKO, for all internal bosonic lines of the Feynman graphs, the following structure generates the effective combined propagator, 
\bea \mathfrak{G}_{\mu \nu}(p)= -ig_{\mu \nu}\left(\frac{1}{p^2}-\frac{1}{p^2-M^2} \right)=-ig_{\mu \nu}\frac{(-M^2)}{p^2(p^2-M^2)}.\eea

\section{The non-relativistic interaction potential}
\label{sec7}
 Following Ref. \cite{Ahluwalia:2022ttu}, here the vertexes, the gauge propagator, and the external ELKO (chosen to be the self-conjugate ones), are employed in the low-energy approach, to obtain the non-relativistic potential. Here, we consider our fundamental model the one composed of a gauge field in the Podolsky sector and a fermionic sector with the generalized dual. First of all, for obtaining the potential, one should use the expression
\bea V(r)=\frac{1}{4E_1E_2}\int e^{i{{\mathbf{p}}\cdot}\mathbf{r}}(-i){\cal{M}}(p)\frac{d^3p}{(2\pi)^3}.\eea
 The associated scattering amplitude reads\footnote{Here the approximation in which the exchanged momentum associated to the propagator line is purely spatial $p^\mu=(0,{\p})$  with a small norm  is employed, such that the particle energy is approximately $\sqrt{m^2+{|\vec q|}^2}$.  Then, we consider that the helicity sums, which are associated to spinor bilinears at the same four-momentum point, are valid up to corrections of order $|{\p}|$.       }
\beq {\cal{M}}_{\upalpha,\upalpha',\upbeta,\upbeta'}(p)&=&i^2\frac{g^2}{m^2}\gdualn\uplambda^S_\upalpha\left(E_q, \vec q-\frac{{\p}}{2}\right)p_{\upgamma^\prime}\sigma^{{\upgamma^\prime}\nu}\uplambda_{\upalpha'}^S\left(E_q,\vec q+\frac{{\p}}{2}\right) \frac{iM^2g_{\nu \mu}}{p^2\left(p^2-M^2 \right)}\nn\\&&\times\gdualn{ \uplambda}^S_\upbeta\left(E_q,\vec q+\frac{{\p}}{2} \right)-p_\omega \sigma^{\omega \mu}\uplambda_{\upbeta'}^S\left(E_q,\vec q -\frac{{\p}}{2}\right).    \eeq
The sign in the momentum of the second vertex is due to the momentum orientation and the Feynman rules as well, in the low energy regime approximation for the spinors.

In order to obtain an expression that do not depend on a specific helicity configuration for external particles, the trace over the helicity space is taken, leading to a potential that is invariant with respect to a change of basis in this internal space\footnote{We have not considered the possibility of a product of traces associated to each vertex since it leads to a vanishing result.}
\begin{align} {\cal{M}}(p)&=\frac{1}{4}\sum_{\upalpha,\upalpha',\upbeta,\upbeta'}\delta^{\upalpha'\upbeta}\delta^{\upalpha \upbeta'}{\cal{M}}_{\upalpha,\upalpha',\upbeta,\upbeta'}(p)=-\frac{1}{4}i^2Tr\left(  p_{\gamma}\sigma^{\gamma\nu} p_\omega \sigma^{\omega}_{\;\, \nu}      \right)\frac{iM^2g^2}{p^2\big(p^2-M^2 \big)}\nonumber \\
&=-3g^2\frac{iM^2}{\big(p^2-M^2 \big).}
\end{align}
 Therefore, at low energies, and for $p^\mu=\big(0,{\p} \big)$ the expression for the potential yields
\bea V(r)=\frac{3g^2}{4m^2}\int \frac{d^3p}{(2\pi)^3}e^{i{{\footnotesize{{\p}}}\cdot\mathbf{r}}}\frac{M^2}{\big({\p}^2+M^2 \big)}=-\frac{3g^2M^2}{4m^2}\frac{e^{-mr}}{4\pi r}.\eea
After obtaining this Yukawa potential, it is a good moment to point out some important features of the model. The derived calculations show a physical system that does not interact at large distances, being dark for cosmology. On the other hand, due to asymptotic freedom, the intensity of the collisions at very small distances approaches zero. Therefore, this system indeed describes dark matter particles in the sense of macro and micro measurements. Using a Maxwell propagator leads to a delta-like contact interaction following the previously mentioned darkness of the phenomenology. 
We must demand that even at cosmological distances, the coupling constant is perturbative. Hence,  as well as ordinary matter, dark matter has a trilinear coupling with photons organizing its structures. The importance of investigating the photon coupling is that it leads to the possibility of measuring dark matter in experiments, due to the existence of effective vertexes associated with the interaction between electrons and also between protons and other ordinary matter particles that interact with light as well. Therefore, considering the vertex studied here, scattering processes with an electron-muon-like qualitative structure  are possible and can be measured. 

Beyond the photon, there is also the interaction with the massive intermediate boson that does not show up as external states. It improves quantum properties and also the behavior of the potential. Moreover, we are going to see that at high momentum, the analogous amplitude for the M{\o}ller scattering goes to zero instead of being constant as for the QED case. It may be interpreted as another factor that makes it hard to detect it in high-energy colliders. Summing up, it gives another contribution to the darkness of such a system.

\section{Corrected M{\o}ller scattering with the generalized ELKO dual and the twisted conjugation prescription}
\label{sec8}
The amplitude for the M{\o}ller scattering, considering the Podolsky propagator and the generalized Elko spinor dual, reads\footnote{We are using the definition  $\chi^\mu(p)=p_{\nu}\sigma^{{\nu}\mu}$.}
\bea
    {\cal{M}}_{\upalpha,\upalpha^\prime,\upbeta,\upbeta^\prime}&=&\frac{M^2g^2}{m^2\big (t-M^2\big)t}i(i)^2\gdualn \uplambda^S_\upalpha(p_4)\chi^\nu(p_1-p_3)\uplambda^S_\upbeta(p_2) \gdualn \uplambda^S_{\upbeta'}(p_3)\chi_{\nu}(p_1-p_3)\uplambda_{\upalpha'}^S(p_1)\nonumber \\
    &&-\frac{M^2g^2}{m^2 \big(u-M^2\big)u}i(i)^2\gdualn \uplambda^S_\upalpha(p_3)\chi^\nu(p_1-p_4)\uplambda^S_{\upbeta}(p_2) \gdualn \uplambda^S_{\upbeta'}(p_4)\chi_{\nu}(p_1-p_4)\uplambda_{\upalpha'}^S(p_1).
\eea
The relative negative sign is due to fermionic exchange of identical electrons in the $u$ channel diagram. The $m$ in the denominator is related to the Feynman rules.  Using the $\ddagger$ twisted conjugation yields
\beq
    \!\!\!\!\!\!\!\!\!\!{\cal{M}}^\ddagger_{\upalpha,\upalpha^\prime,\upbeta,\upbeta^\prime}&=&\frac{M^2g^2}{m^2\big(t-M^2\big)t}(-)i(i)^2\gdualn \uplambda^S_{\upalpha'}(p_1)\chi_\nu(p_1-p_3)\uplambda^S_{\upbeta'}(p_3)\gdualn \uplambda^S_{\upbeta}(p_2)\chi^{\nu}(p_1-p_3)\uplambda_\upalpha^S(p_4)\nn \\
    &&-\frac{M^2g^2}{m^2\big(u-M^2\big)u}(-)i(i)^2\gdualn \uplambda^S_{\upalpha'}(p_1)\chi^\nu(p_1-p_4)\uplambda_{\upbeta'}^S(p_4)\gdualn \uplambda^S_\upbeta(p_2)\chi_{\nu}(p_1-p_4)\uplambda^S_\upalpha(p_3). 
\eeq
Performing the sum over the external particle helicities implies that  
\bea
    \sum_{\upalpha,\upalpha',\upbeta,\upbeta'}\!\!\!\!{\cal{M}}_{\upalpha,\upalpha^\prime,\upbeta,\upbeta^\prime}{\cal{M}}^\ddagger_{\upalpha,\upalpha^\prime,\upbeta,\upbeta^\prime}\!=\!\frac{M^4g^4}{\big(t-M^2\big)^2t^2}Tr\big\{\chi^\gamma(p_1-p_3)\chi^{\nu}(p_1-p_3)\big\}Tr\big\{\chi_\nu(p_1-p_3)\chi_{\gamma}(p_1-p_3)\big\}\nonumber \\
    +\frac{M^4g^4}{\big(u-M^2\big)^2u^2}Tr\big\{ \chi^\gamma(p_1-p_4)\chi^{\nu}(p_1-p_4)\big \}Tr\big\{\chi_\nu(p_1-p_4)\chi_{\gamma}(p_1-p_4)\big\}\nonumber \\
   \!\!\!\!\!\!\! \!\!\!\!\!\!\! \!\!\!\!\!\!\! \!\!\!\!\!\!\! -\frac{2M^4g^4}{\big(u\!-\!M^2\big)u\big(t\!-\!M^2\big)t}Tr\big\{ \chi^\gamma(p_1\!-\!p_3)\chi^{\nu}(p_1\!-\!p_3)\chi_\gamma(p_1\!-\!p_4)\chi_{\nu}(p_1\!-\!p_4)\big\}.\nonumber
\eea
Computing the traces, using  $Tr\big\{\chi_\nu(p)\chi_{\mu}(p)\big\}=-4p^2\theta_{\mu\nu}$ and $\theta^{\mu \nu}\theta_{\mu \nu}=3$, and averaging, yields
\beq
   \frac{1}{4}\sum_{\upalpha,\upalpha',\upbeta,\upbeta'}{\cal{M}}_{\upalpha,\upalpha^\prime,\upbeta,\upbeta^\prime}{\cal{M}}^\ddagger_{\upalpha,\upalpha^\prime,\upbeta,\upbeta^\prime}&=&\frac{12M^4g^4t^2}{\big(t-M^2\big)^2t^2}
    +\frac{12M^4g^4u^2}{\big(u-M^2\big)^2u^2}
    +6\frac{M^4g^4\big((p_1-p_4)_\mu(p_1-p_3)^\mu \big)^2}{\big(u-M^2\big)u\big(t-M^2\big)t}\nonumber\\
    &=&\frac{12M^4g^4}{\big(t-M^2\big)^2}
    +\frac{12M^4g^4}{\big(u-M^2\big)^2}.
\eeq
 The crossed term vanishes. It can be easily understood considering the Mandelstam variables. This amplitude vanishes at high energies and tends to be a constant in the low energy regime.
It is worth emphasizing that  light ELKO coupling, which
might evade any collider constraint do not need to be excluded, and the phenomenological necessity, demanding that the M{\o}ller scattering must remain unitary up to 8 TeV \cite{Alves:2017joy}, can be reformulated  considering our improvements.

\section{ELKO pair annihilation }
\label{sec9}
 In order to evaluate the relic density and the associated freeze out temperature, important phenomenological content to constrain the model, the amplitude for the ELKO pair annihilation must be computed. One can use the approach in Ref. \cite{Agarwal:2014oaa}, however  now using the correct and unitary prescription for the generalized spinor dual and for the twisted conjugation.  The same model employed in the previous section will be regarded. Then, considering the expected value for this temperature due to cosmology investigations for dark matter, it is possible to find constraints for the coupling $g$ and the ELKO mass in a future work. The relevant amplitude has contributions from the $t$ and $u$ channels. It reads\footnote{The definition $\chi^\alpha(p)=p_{\beta}\sigma^{{\beta}\alpha}$ is used.}
\bea
    {\cal{M}}_{\upalpha,\upalpha^\prime,\upbeta,\upbeta^\prime}&=&\frac{g^2}{m\big(t-m^2\big)}i(i)^2\gdualn \uplambda^A_\upbeta(p_2)\chi^\nu(p_1-p_3)\chi^{\sigma}(p_1-p_3)\epsilon_\nu^\upalpha(p_4) \epsilon^{\upalpha'}_\sigma(p_3)\uplambda_{\upbeta'}^S(p_1)\nonumber \\
    &&+\frac{g^2}{m\big(u-m^2\big)}i(i)^2\gdualn \uplambda^A_\upbeta(p_2)\chi^\sigma(p_1-p_4)\chi^{\nu}(p_1-p_4)\epsilon_\nu^{\upalpha}(p_4) \epsilon^{\upalpha'}_\sigma(p_3)\uplambda_{\upbeta'}^S(p_1).
\eea
The relative positive sign is due to bosonic photon exchange in the $u$ channel diagram. The $m$ in the denominator is related to the Feynman rules. Using the $\ddagger$ twisted conjugation implies that 
\bea
    {\cal{M}}_{\upalpha,\upalpha^\prime,\upbeta,\upbeta^\prime}^\ddagger&=&\frac{g^2}{m\big(t-m^2\big)}(-)i(i)^2\gdualn \uplambda^A_{\upbeta'}(p_1)\chi^\upsigma(p_1-p_3)\chi^{\nu}(p_1-p_3)\epsilon_\nu^{*\upalpha}(p_4) \epsilon^{*\upalpha'}_\upsigma(p_3)\uplambda_{\upbeta}^S(p_2)\nonumber \\
    &&+\frac{g^2}{m\big(u-m^2\big)}(-)i(i)^2\gdualn \uplambda^A_{\upbeta'}(p_1)\chi^\nu(p_1-p_4)\chi^{\sigma}(p_1-p_4)\epsilon_\nu^{*\upalpha}(p_4) \epsilon^{*\upalpha'}_\sigma(p_3)\uplambda_{\upbeta}^S(p_2).
\eea 
Then, summing over the helicities over the product of these terms yields
\bea
    \!\!\!\!-\!\!\!\!\!\sum_{\upalpha,\upalpha^\prime,\upbeta,\upbeta^\prime}\!\!\!{\cal{M}}_{\upalpha,\upalpha^\prime,\upbeta,\upbeta^\prime}{\cal{M}}^\ddagger_{\upalpha,\upalpha^\prime,\upbeta,\upbeta^\prime}=\frac{g^4}{\big(t-m^2\big)^2}Tr\big\{\chi^\upsigma(p_1-p_3)\chi^{\nu}(p_1-p_3)\chi_\nu(p_1-p_3)\chi_{\upsigma}(p_1-p_3)\big\}\nonumber \\
    +\frac{g^4}{\big(u-m^2\big)^2}Tr\big\{ \chi^\upsigma(p_1-p_4)\chi^{\nu}(p_1-p_4)\chi_\nu(p_1-p_4)\chi_{\upsigma}(p_1-p_4)\big\}\nonumber \\
    +\frac{2g^4}{\big(u-m^2\big)\big(t-m^2\big)}Tr\big\{ \chi^\upsigma(p_1-p_3)\chi^{\nu}(p_1-p_3)\chi_\upsigma(p_1-p_4)\chi_{\nu}(p_1-p_4)\big\}.
\eea
This extra overall minus sign is due to the latter polarization sum. Therefore, as previously discussed, it can be associated with probability density, including an averaging factor, as \beq
P=\frac{1}{4}\sum_{\upalpha,\upalpha^\prime,\upbeta,\upbeta^\prime}{\cal{M}}_{\upalpha,\upalpha^\prime,\upbeta,\upbeta^\prime}{\cal{M}}_{\upalpha,\upalpha^\prime,\upbeta,\upbeta^\prime}^\ddagger (-1)^N(-1)^V,
\eeq with $V=2$ and $N=1$, eliminating the minus sign.  Therefore it yields 
\bea
    \!\!\!\!-\frac{1}{4}\sum_{\upalpha,\upalpha^\prime,\upbeta,\upbeta^\prime}{\cal{M}}_{\upalpha,\upalpha^\prime,\upbeta,\upbeta^\prime}{\cal{M}}_{\upalpha,\upalpha^\prime,\upbeta,\upbeta^\prime}^\ddagger=\frac{9g^4t^2}{\big(t-m^2\big)^2}
    +\frac{9g^4u^2}{\big(u-m^2\big)^2}
    -6\frac{g^4\big((p_1-p_4)_\mu(p_1-p_3)^\mu \big)^2}{\big(u-m^2\big)\big(t-m^2\big)}.
\eea
Considering the definition of Mandelstan variables we have $(p_1-p_4)_\mu(p_1-p_3)^\mu=0$. Then, at low energies, since $t$ and $u$ are proportional to ${\p}^2$, 
\bea
    \frac{1}{4}\sum_{\upalpha,\upalpha^\prime,\upbeta,\upbeta^\prime}{\cal{M}}_{\upalpha,\upalpha^\prime,\upbeta,\upbeta^\prime}{\cal{M}}_{\upalpha,\upalpha^\prime,\upbeta,\upbeta^\prime}^\ddagger \sim
  0.
\eea

\section{ELKO-proton scattering, an effective vertex}
\label{snew}
 In order to detect dark matter in laboratory, it must interact with ordinary matter associated to the detector devices. Therefore, by means of the ELKO-photon interaction vertex, a M${\o}$ller like scattering between dark matter and protons can occur and, at principle, be measured.\\
\indent For practical calculation, one needs the effective vertex describing the interaction of this fermionic particle with light. At the low transferred energy regime, it reads
\bea \mathcal{V}_\mu\equiv \frac{1}{m'}\sigma_{\mu \nu}p^\nu+\gamma_\mu, \eea
with $m'$ being the proton mass.  The expression for this amplitude is given by
\bea
    {\cal{M}}_{\upalpha,\upalpha^\prime,\upbeta,\upbeta^\prime}&=&\frac{M^2eg}{m\big (t-M^2\big)t}i(i)^2\gdualn \uplambda^S_\upalpha(p_4)\chi^\nu(p_1-p_3)\uplambda_\upbeta^S(p_2) \bar u_{\upbeta'}(p_3)\mathcal{V}_{\nu}(p_1-p_3)u_{\upalpha'}(p_1),
    \eea
    whose twisted  conjugation yields
    \beq
    \!\!\!\!\!\!\!\!\!\!{\cal{M}}^\ddagger_{\upalpha,\upalpha^\prime,\upbeta,\upbeta^\prime}&=&\frac{M^2eg}{m\big(t-M^2\big)t}(-)i(i)^2\bar u_{\upalpha'}(p_1)\mathcal{V}_\nu^{'}(p_1-p_3)u_{\upbeta'}(p_3)\gdualn \uplambda^S_{\upbeta}(p_2)\chi^{\nu}(p_1-p_3)\uplambda_\upalpha^S(p_4),\nn 
\eeq 
with $\mathcal{V}_\mu^{'} \equiv -\frac{1}{m'}\sigma_{\mu \nu}p^\nu+\gamma_\mu$ and $u(p)$ denoting a spinor that represents an incoming proton. 
Then, summing over the helicities yields
\beq&&\sum_{\upalpha,\upalpha',\upbeta,\upbeta'}{\cal{M}}_{\upalpha,\upalpha^\prime,\upbeta,\upbeta^\prime}{\cal{M}}^\ddagger_{\upalpha,\upalpha^\prime,\upbeta,\upbeta^\prime}\nn\\
&=&\frac{M^4g^2e^2}{\big(t-M^2\big)^2t^2}Tr\big\{\chi^\gamma(p_1-p_3)\chi^{\nu}(p_1-p_3)\big\}\times Tr\big\{\mathcal{V}_\nu(p_1-p_3)\big(\slashed{p}_1+m'\big)\mathcal{V}^{'}_{\gamma}(p_1-p_3)\big(\slashed{p}_3+m'\big)\big\}\nn\\&=&\frac{M^4g^2e^2}{\big(t-M^2\big)^2t^2}\left[-16t^2+8t\left(4{m'}^2 +\frac{t^2}{{m'}^2}    \right)\right].
\eeq
Regarding this amplitude, $V=1$ and $N_f=0$. Then, according to the established rule, averaging over the helicity sum, the following positive definite probability density is associated to the physical process
\bea P=(-1)^0(-1)^1\frac{1}{4}\sum_{\upalpha,\upalpha',\upbeta,\upbeta'}{\cal{M}}_{\upalpha,\upalpha^\prime,\upbeta,\upbeta^\prime}{\cal{M}}^\ddagger_{\upalpha,\upalpha^\prime,\upbeta,\upbeta^\prime}=\left[4t^2-2t\left(4{m'}^2 +\frac{t^2}{{m'}^2}    \right)\right]\frac{M^4g^2e^2}{\left(t-M^2\right)^2t^2}.\eea

The XENON1T experiment has the goal of directly  detecting dark matter in the laboratory as weakly interacting particles \cite{XENON:2019ykp}. 
One can therefore use the M{\o}ller and the pair annihilation, together with the ELKO-proton scattering studied in this section, to obtain experimental limits for the model parameters via XENON1T-type experimental cross sections and through freeze-out temperature,  using the structures obtained in these last three sections.

\section{Linear term from the Dirac dual and a solution exhibiting a galaxy flat rotation curve}
\label{sec10}
In this section, we are going to investigate the fermionic response associated with the exotic theory constructed with the Dirac and not the generalized  ELKO dual (\ref{eq116}) with the addition of a Podolskyan propagator. It is associated with the possibility of a solution in which the so-called flat rotation curve for a planar galaxy occurs. These are consequences of the renormalized ELKO coupled system of equations of motion that, at low energies, due to the presence of the radiative linear term from the self-energy\footnote{See Section \ref{sec4}.}, reads
\bea \big(iA \gamma^\mu \partial_\mu+ \Box+m^2\big)\Uppsi(\vec r,t)=0 \eea
   with $A$ fixed by the mentioned extra renormalization condition, whose associated experimental input can be of cosmological nature.    This kind of differential operator has a Chern number associated with it, which is a functional of this operator in the momentum space calculated in the frame $p^\mu=(0,{\p})$. In the context of condensed matter, namely for topological insulators, it leads to interesting localized states on the boundary of the sample \cite{Monastyrsky:1984ez,Schindler:2020crr}. At principle, our goal here would be to emulate it in the context heretofore presented. The mentioned Chern number reads 
\bea  n_{{\scalebox{0.6}{$\textsc{Chern}$}}}=\frac{1}{2}\left[{{\scalebox{0.9}{$\textsc{sign}$}}}\left(-\frac{1}{A}\right)+{{\scalebox{0.9}{$\textsc{sign}$}}}\left(\frac{m^2}{A}\right) \right],  \eea
which is non-trivial just for tachyonic particles, which is not our case. Despite this fact, since there are linear and quadratic terms in derivatives, considering large distances and low energy, a fast decaying solution displaying interesting properties can be possibly derived. Considering spherical coordinates  with the fixed azimuthal angle $\theta=\frac{\pi}{2}$, to study planar galaxies in form of a disc, and discarding derivative terms with $1/r$ factor due to the cosmological distances, a specific solution is given by 
\bea \big[iA \gamma^r (\partial_r e^{\Omega r})\Uplambda(\vec r,t)-(\partial^2_re^{\Omega r})\Uplambda(\vec r,t)-Am\Uppsi(\vec r,t)\big]=0,\label{cont}  \eea
with $\gamma^r = \gamma^1\sin\theta\cos\phi+\gamma^2\sin\theta\sin\phi+\gamma^3\cos\theta$. The ansatz  for the conposed ELKO is explicitly given by 
\beq
\Uppsi(\vec r,t)=\Uplambda(\vec r,t)e^{\Omega r},\eeq
with $\Uplambda(\vec r,t)$ being a linear combination of ELKO self-conjugate fermions with different helicities, 
\bea  \Uplambda(p^\mu)=\uplambda^S_-(p^\mu)-i\uplambda_+^S(p^\mu).   \label{elkol}            \eea
Taking into account the coupled system \eqref{coupled} that consists of the equations of motion for the types of ELKO, therefore 
the equation of motion for the composed ELKO (\ref{elkol}) reads  \beq\slashed{p}\Uplambda(p^\mu)=m\Uplambda(p^\mu).\eeq
This condition also implies that $\Uplambda(\vec r,t)$ is an element in the kernel of the Klein--Gordon operator, $(\Box+m^2)\mathbb{I}$. The defining equation (\ref{cont}) was obtained by means of the use of these properties.
 The explicit matrix form for this equation is given below 
 \begin{eqnarray}
   {\mathcal{K}}_{4 \times 4}\Uppsi(\vec r,t)=0,
\end{eqnarray}    
where
\beq
{\mathcal{K}}_{4 \times 4}\equiv \left[\begin{array}{cccc}
                                 
                                 -G & 0 & 0 & iAe^{-i\phi}\Omega \\
                                 0 & -G & iAe^{i\phi}\Omega & 0   \\
                                 0 & -iAe^{-i\phi}\Omega & -G& 0  \\
                                 -iAe^{i\phi}\Omega &  0 & 0&-G \\
                               \end{array}\right]=-G\mathbb{I}-A\Omega (\sigma_3\otimes\sigma_3)\mathfrak{G}(p^\mu),
\eeq
with $G\equiv \Omega^2+Am$ and $\sigma_3=\scriptsize{\begin{pmatrix}1&0\\0&-1\end{pmatrix}}$ is the spin Pauli matrix. It is worth to mention the central role played by the  operator $\mathfrak{G}(p^\mu)$ in Eq. (\ref{op}), 
entering the ELKO spin sums (\ref{ss1}) and with the generalized ELKO dual, in Eqs. (\ref{eq116}, \ref{ss3}). 
The freedom in fixing a counterterm for the linear term is used to eliminate divergences as fixing $A$ with a renormalization condition at the low-energy regime, also demanding that $A<0$, with an extremely small modulus to reproduce the qualitative aspect of the flat rotation curve.  The factor $\Omega$ can be found by demanding a vanishing determinant as a necessary condition for this equation to hold. After it, the correspondent ELKO solution can be obtained. 
 To derive $\Omega$, the following identity may be employed,
\bea {\mathcal{K}}_{4 \times 4}=\begin{pmatrix}
\tilde A_{2\times2}	&  B_{2 \times 2}  \\
	C_{2\times 2}	&  D_{2\times 2}\end{pmatrix},\quad \quad \det {\cal{K}}=\det D \det(\tilde A-BD^{-1}C).\eea 
Particularizing for the case here studied  yields 
\bea  \det {\cal{K}}=\det G^2\det\begin{pmatrix}
G-G^{-1}\Omega^2 A^2	&  0 \\
	0	&  G-G^{-1}\Omega^2 A^2  
	\end{pmatrix}.     \eea 
Therefore, there are four possibilities given by
\beq
\Omega=\begin{cases}&\pm \sqrt{|A|m},\\
&\frac12\left({\mp |A|\pm \sqrt{A^2-4mA}}\right)\sim \pm \frac{A}{2}\pm i\sqrt{m|A|}.
\end{cases}
\eeq
The solution can be therefore split into two parts. The first one lies in the range starting from the center of the galaxy up to a critical radius $r_c$. The second range  starts from this point. The  disc with radius $r_c$ corresponds to the dark matter halo. Therefore, this halo do not disappear far from $r_c$, this is just the region of maximal dark matter density. In the region $I$ it is increasing and, in region $II$, it decreases. A possible solution for the first range,  $0<r<r_c$, is given by
\bea \Uppsi(\vec r,t)=\left(i\tilde Ae^{\frac{|A|}{2}(r-r_c)}+Be^{-\frac{|A|}{2}(r-r_c)}\right) \exp\left[{i\sqrt{|A|m}(r-r_c)}\right]\Uplambda(\vec r,t).    \eea
It is important to note that
\bea  \Uplambda^\dagger (p^\mu)\gamma_0\Uplambda(p^\mu)=4m,\eea
 yielding 
\bea    \frac{4m}{(2\pi)^4}\int d^4p=\int d^4x \bar \Uplambda(x)  \Uplambda(x).              \eea 
Hence,
\bea   \frac{4m V_p}{(2\pi)^4V} = \bar \Uplambda(x)  \Uplambda(x),     \eea
with $V_p$ being the volume of the 4-dimensional momentum space and $V$ is the  spacetime volume. Considering our solution for cosmological distances,  spatial momentum and, therefore, the spatial derivatives ($ \hat p_k=-i\partial_k$) can be can neglected. The radial one leads just to a  contribution of order $\sim |A|$, when acting on the exponential part, and another tiny contribution from the $\Uplambda(x)$ structure at large cosmological distances, corresponding to the low momentum regime. The same occurring for the remaining spatial derivatives. In order to derive an expression for the matter density, one must consider these latter approximations in the $T_{00}$ component of the energy momentum tensor of Ref. \cite{Ahluwalia:2022ttu}. Only the scalar-like part of the tensor will be employed, since the fermionic-like sector depends on the four-momentum of the graviton,  which is massless and, at low energies, can be discarded. The scalar sector, for the case of an approximately flat geometry, reads
\bea T_{00}^{{\scalebox{0.6}{$\textsc{scalar}$}}}(x)=\frac{1}{2}\partial_0 \bar\Uppsi(x)\partial_0 \Uppsi(x)-\frac{g_{00}}{2}(\partial_i \bar\Uppsi(x)\partial_i \Uppsi(x)-m^2\bar\Uppsi(x)\Uppsi(x)).           \eea
we have discarded the contributions from the spatial derivatives owing to the previously exposed reasons.

 Using the equation of motion of the composed ELKO, $\slashed{\partial}\Uplambda(x)=im\Uplambda$, we can derive, for the approximation employed here $\slashed{p} \mapsto \gamma_0p_0$, the following results
\bea \partial_0\Uplambda(x)=im\gamma_0\Uplambda(x), \qquad \partial_0\bar \Uplambda(x)=-im\bar \Uplambda(x) \gamma_0.\eea The energy density, in this approximation,  reads
\bea T_{00}(x)=\frac{4m^3V_p}{(2\pi)^4V}|\tilde A|^2e^{|A|(r-r_c)}+\frac{4m^3V_p}{(2\pi)^4V}|B|^2e^{-|A|(r-r_c)}. \eea
 It is necessary to conveniently choose constants to properly identify this with a matter density.
Besides, one can verify that 
\bea \bar\Uppsi(\vec r,t)\Uppsi(\vec r,t)=\frac{4mV_p}{(2\pi)^4V}|\tilde A|^2e^{|A|(r-r_c)}+\frac{4mV_p}{(2\pi)^4V}|B|^2e^{-|A|(r-r_c)}. \eea Considering a  galaxy in the form of a disc, one can obtain\footnote{In this integration, we considered the term $e^{|A|(R-r_c)}\sim 1$ in the subtraction on the integration limits.}
\bea M_I = 2\pi \int_0^R r\,T_{00}(r)\,dr,  \label{mass}  \eea
yielding
\bea M_I \sim \frac{4m^3V_p}{(2\pi)^4V}\frac{32\pi}{|A|} \left[|\tilde A|^2e^{|A|(R-r_c)}   -|B|^2e^{-|A|(R-r_c)}\right]R,  \eea
where one must demand $|\tilde A|^2>|B|^2$. The following range of model parameters have been considered  $|A(R-r_c)|\lesssim10^{-1}$ and also  $|A(r_c)|\lesssim10^{-1}$. Then, it leads to a constraint for experimental limits regarding $A$ in order to derive a galaxy dark matter mass that grows approximately \footnote{In fact, without considering any approximation, it is proportional to $\left[\frac{|\tilde A|^2}{|A|}e^{|A|(R-r_c)}\big(R+\frac{1-e^{-|A|R}}{|A|} \big)  -\frac{|B|^2}{|A|}e^{-|A|(R-r_c)}\big(R+\frac{1-e^{|A|R}}{|A|} \big)\right]$.} at a linear rate, with $e^{-|A|(R-r_c)}\sim 1$ and  $e^{|A|(r_c)}\sim 1$. Therefore, if one wants to reproduce cosmological data, the intensity of the coefficient of the anomalous self energy linear term must be fixed in order to fulfil this condition. Then, the order of magnitude for the halo $r_c$ and the visible galaxy radius should be taken into account since, for region $I$, $R_g<R<r_c$. This procedure can be understood as a renormalization condition for this anomalous self energy contribution.

 Considering these constraints and denoting by $R_g$ the visible galaxy radius, the velocity for $R_g<R<r_c$ the region in which a unexpected flat rotation curve occurs
\bea  v^2(R)=\frac{G}{R}\left[M_T+\frac{8m^3V_p}{|A|\pi^3V} \left(|\tilde A|^2   -|B|^2\right)R\right],        \eea
where $M_T$ denotes the total visible mass. 
The constants involved in the previous equations can be fixed by using  experimental and observational links to cosmology, as flat rotation curves. Limits can be hence imposed for the constant coupling through this type of data. Ref. \cite{Pereira:2018xyl} imposes experimental limits on the ELKO mass  through thermodynamic considerations related to the order of magnitude of the size of a galaxy and the degeneracy pressure to maintain a volume of free ELKO gas for the case of these dimensions. It would be interesting to develop a similar approach for our planar disc galaxy case.\\
\indent For completeness, we present the solution for the region outside the halo $R>r_c$ with decreasing dark matter density 
\bea \Upsigma(\vec r,t)=\left(i Ce^{-\frac{|A|}{2}(r-r_c)}e^{\left[{i\sqrt{Am}(r-r_c)}\right]}+De^{-\frac{|A|}{2}(r-r_c)}e^{\left[{-i\sqrt{|A|m}(r-r_c)}\right]}\right) \Uplambda(\vec r,t)    \eea
To have a well-defined solution,  continuity at $r_c$ must be demanded, 
\bea  \Upsigma(\vec r_c,t)=\Uppsi(\vec r_c,t), \qquad\qquad \quad \frac{\partial}{\partial r}\Upsigma(\vec r,t)\vert_{r_c}=\frac{\partial}{\partial r}\Uppsi(\vec r,t)\vert_{r_c}.     \eea
 These conditions lead to\footnote{For these real parameters, the equalities $\tilde A=C$ and $D=B$ hold.} $i\tilde A+B=iC+D$ and $\tilde A=-\frac{ 2B\sqrt{|A|m}}{|A|}$, respectively. As previously mentioned, physical solutions must have $|\tilde A|>|B|$. Finally, it is important to mention that, for region $II$, we have $R>r_c$. Then, for sufficiently large distances $ |A(R-r_c)| \geq 1$, the exact  result for the integration (\ref{mass}) must be considered,  leading to  an exponential decaying behaviour describing the expected continuously decreasing mass profile in $II$.
 
\section{Concluding remarks and perspectives}
\label{sec11}
We investigated the completely peculiar role played by ELKO in QFT as non-standard quantum fields with new physical signatures. For the ELKO-photon interaction, the spin sums and the Feynman rules were derived, using the Dirac spinor dual. Due to the use of the Dirac spinor dual, some issues  involving the polarization tensor and
bubble diagrams of ELKO and anti-ELKO are presented and solved. Also, unitarity aspects of ELKO in this context are discussed. The Podolsky sector in a QFT for ELKO was introduced and the associated propagator explored, together with radiative corrections. Renormalization features
were also addressed. When taking into account the ELKO generalized spinor dual, instead of the Dirac spinor dual, the unitarity was analyzed and
the different spinor field conjugations were analyzed, with the use of the twisted conjugation. Therefore, the polarization tensor regarding
ELKO was computed, regulating divergences under renormalization procedures.  ELKO self-interaction was then included and decay rates were  studied, also
employing the Podolsky propagator. The associated beta functions have been also computed. With these tools, one can refine constraints regarding the dark matter described by ELKO in monophoton events at the LHC. Podolsky QFT and unitarity aspects involving ELKO scattering processes were addressed, where the vertexes, the gauge propagator, and the
external ELKO were scrutinized in the low-energy approach, to obtain the non-relativistic potential.
To study the scattering amplitude, 
the Podolsky gauge sector field and  ELKO with the generalized dual were employed, yielding Yukawa-like potential. The interaction between the ELKO and the proton was also investigated.   The corrected M{\o}ller scattering with the generalized ELKO
spinor dual as well as the twisted conjugation was discussed. We also derived an important result for a future investigation of the relic density and the freeze-out temperature, computing the scattering amplitude for the
ELKO pair annihilation. When using a linear term arising from the self-energy
corrections for the model composed of ELKO spinors with the Dirac dual, a 
solution exhibiting a galaxy flat rotation curve was studied,  implementing the ELKO fermionic
response associated with the exotic theory constructed with the Dirac spinor dual instead of the
generalized ELKO dual, with the addition of a Podolskyan propagator. The results in the context
of a comparison with  topological insulators tools in condensed matter were also discussed. Besides, we proved that the Lagrangian  (\ref{ll1}), which carries the interaction between ELKO and the electromagnetic strength tensor, is invariant under the twisted conjugation.

As ELKO is a dark quantum field, being a prime candidate to describe dark matter particles, the tree level ELKO-Higgs interaction can be also further studied in the context here presented, emulating the previous results in Ref. \cite{Alves:2014kta}, wherein ELKO yields the  relic abundance probed by  WMAP. Also, a triple coupling scenario can be scrutinized, for the ELKO  mass
can be generated by a mechanism related to the electroweak symmetry breaking.
The Higgs decay into ELKO/anti-ELKO pair, with monojet production, can be also better studied \cite{Alves:2014qua}.
\bigskip
\bigskip\bigskip\bigskip
\paragraph*{Declaration of competing interest.} The authors declare that they have no known competing financial interests or personal relationships that could have appeared to influence the work reported in this paper.

\paragraph*{Data Availability Statements:}  the datasets generated during and/or analyzed during the current study are available from the corresponding author upon reasonable request.

\subsubsection*{Acknowledgement}
GBG thanks to The São Paulo Research Foundation -- FAPESP  Post Doctoral grant No. 2021/12126-5.
RdR~is grateful to FAPESP (Grants No. 2021/01089-1 and No. 2022/01734-7) and the National Council for Scientific and Technological Development -- CNPq (Grants No. 303390/2019-0 and No. 406134/2018-9), for partial financial support.

\bibliography{elko.bib}

\end{document}